\documentclass[11pt,a4paper]{article}
\usepackage{jheppub}
\usepackage{slashed}
\usepackage[utf8]{inputenc}
\usepackage{amsthm}
\newtheorem{theorem}{Theorem}

\theoremstyle{definition}
\newtheorem*{definition}{Definition}

\usepackage{bm} 
\usepackage{accents}

\newcommand{\wad}[2]{\widehat{a}^\dagger_#2 (#1)}
\newcommand{\wa}[2]{\widehat{a}_#2 (#1)}
\newcommand{\wbd}[2]{\widehat{b}^\dagger_#2 (#1)}
\newcommand{\wb}[2]{\widehat{b}_#2 (#1)}
\newcommand{\di}{{\rm d}}
\newcommand{\dist}{{\rm dist}}
\newcommand{\ii}{i}
\newcommand{\unt}[1]{\underset{\smile}{#1}}
\newcommand{\upt}[1]{\stackrel{\frown}{#1}}

\def\wT{{\widehat T}}
\def\wj{{\widehat j}}
\def\wQ{{\widehat Q}}
\def\wA{{\widehat A}}
\def\wB{{\widehat B}}
\def\wP{{\widehat P}}
\def\wJ{{\widehat J}}

\def\wrho{{\widehat{\rho}}}

\def\codevnu{{\stackrel{\leftrightarrow}{\partial^\nu}}}
\newcommand{\tr}{{\rm tr}}  
\newcommand{\Tr}{{\rm Tr}}  
  
\newcommand{\e}{{\rm e}}

\newcommand{\omegav}{\boldsymbol{\omega}}

\newcommand{\p}{{\rm p}}

\newcommand{\subs}[1]{_{\textup{#1}}}

\newcommand{\be}{\begin{equation}}
\newcommand{\ee}{\end{equation}}                                                                               
\newcommand{\bea}{\begin{eqnarray}}
\newcommand{\eea}{\end{eqnarray}}

\begin{document}

\title{Exact equilibrium distributions in statistical quantum field theory with rotation 
and acceleration: Dirac field}
\author[a,b]{A. Palermo,}
\author[a,c]{M. Buzzegoli}
\author[a]{and F. Becattini}
\affiliation[a]{Dipartimento di Fisica e Astronomia, Universit\`a di Firenze and INFN Sezione di Firenze,\\
Via G. Sansone 1, Sesto Fiorentino I-50019, Firenze, Italy}
\affiliation[b]{Institut für Theoretische Physik, Johann Wolfgang Goethe-Universität, Max-von-Laue-Str. 1, 60438 Frankfurt am Main, Germany}
\affiliation[c]{Department of Physics and Astronomy, Iowa State University, Ames, Iowa 50011, USA}
\emailAdd{andrea.palermo@unifi.it}
\emailAdd{matteo.buzzegoli@unifi.it}
\emailAdd{becattini@fi.infn.it}

\abstract{
We derive the general exact forms of the Wigner function, of mean values of conserved currents,
of the spin density matrix, of the spin polarization vector and of the distribution function of massless
particles for the free Dirac field at global thermodynamic equilibrium with rotation and
acceleration, extending our previous results obtained for the scalar field. The solutions are obtained 
by means of an iterative method and analytic continuation, which lead to formal series in thermal 
vorticity. In order to obtain finite values, we extend to the fermionic case the method of analytic
distillation introduced for bosonic series. The obtained mean values of the stress-energy tensor, vector 
and axial currents for the massless Dirac field are in agreement with known analytic results in the special
cases of pure acceleration and pure rotation. By using this approach, we obtain new expressions of
the currents for the more general case of combined rotation and acceleration and, in the pure acceleration
case, we demonstrate that they must vanish at the Unruh temperature. 
}

\maketitle
\clearpage
\section{Introduction}

In a previous paper~\cite{ScalarField}, we presented a new method to calculate exact expressions
of mean values involving the free scalar field in the most general global thermodynamic equilibrium 
in Minkowski space-time, including acceleration and rotation. In this paper, we extend that method to a free 
quantum field of any spin and particularly the free Dirac field of spin $1/2$ particles. We obtain exact 
expressions of several physical quantities, including the covariant Wigner function, which is of great
relevance for the relativistic kinetic theory of massive fermions \cite{Weickgenannt:2019dks,Weickgenannt:2020aaf,Wang:2019moi,Hattori:2019ahi,Gao:2019znl,Wang:2020pej,Yang:2020hri} 
as well as the spin density matrix and the spin polarization vector, which are of special interest for the 
phenomenology of relativistic heavy ion collisions \cite{Becattini:2020ngo}.
We demonstrate the viability of our method by comparing the newly found results with those known in the 
literature for the two special cases of equilibrium with pure acceleration and pure rotation and with
those known at the leading perturbative order in thermal vorticity. On top of that, we compute the exact stress 
energy tensor and axial current for a massless field at equilibrium with both acceleration and rotation.
Furthermore, we derive the exact expression of the massless fermions distribution function at global 
equilibrium, which is very relevant for the chiral kinetic theory \cite{Son:2012zy,Stephanov:2012ki,Chen:2014cla,Hidaka:2016yjf,Gao:2017gfq,Shi:2020htn,Liu:2018xip,Manuel:2013zaa,Huang:2018wdl,PhysRevLett.110.262301,PhysRevD.97.051901,Liu:2020flb} as well as the 
exact formula of the spin polarization vector for massive particles as a formal series which extends 
the known expression at first order.

The density operator describing the most general thermodynamic equilibrium in flat space-time 
reads~\cite{Becattini:2012tc}:
\be\label{general}
  \wrho = \frac{1}{Z} \exp \left[ - b_\mu {\wP}^\mu  
  + \frac{1}{2} \varpi_{\mu\nu} \wJ^{\mu\nu} + \zeta \wQ \right],
\ee
where $\wP^\mu$ and $\wJ^{\mu\nu}$ are the generators of translation and Lorentz transformations 
and $\wQ$ is a conserved charge; the four-vector $b$ is constant and time-like, $\varpi$ is an 
anti-symmetric constant tensor, the {\em thermal vorticity}, and $\zeta$ is the ratio between the 
chemical potential and the temperature, i.e. $\zeta=\mu/T$. This operator and its characteristics 
have been studied in detail elsewhere \cite{Becattini:2015nva,Buzzegoli:2018wpy,Becattini:2014yxa,Becattini:2019dxo}.
We just remind that the local temperature $T$ measured by a comoving thermometer is $1/\sqrt{\beta^2}$, 
where $\beta$ is the four-temperature vector. For the density operator \eqref{general} it is 
the Killing vector:
\be\label{fourtemp}
 \beta_\mu = b_\mu + \varpi_{\mu\nu} x^\nu,
\ee
and its direction defines the flow velocity $u$ in space-time \cite{Becattini:2014yxa}. The thermal 
vorticity tensor can be decomposed as: 
\begin{equation*}
  \varpi_{\mu\nu} = \epsilon_{\mu\nu\rho\sigma} w^\rho u^\sigma +
  	(\alpha_\mu u_\nu - \alpha_\nu u_\mu),
\end{equation*}
with:
\be\label{alphaw}
  \alpha^\mu = \varpi^{\mu\nu} u_\nu \qquad \qquad 
   w^\mu =-\frac{1}{2}\epsilon^{\mu\nu\rho\sigma}\varpi_{\nu\rho} u_\sigma
\ee
and, at the global equilibrium defined by the equation \eqref{fourtemp}, $w$ and $\alpha$ are related to 
the acceleration and kinematic vorticity fields $A^\mu$ and $\omega^\mu$, by the simple relations:
$$
  \alpha^\mu = \frac{A^\mu}{T},  \qquad \qquad w^\mu =\frac{\omega^\mu}{T}.
$$
It is important to stress that, even though the above definitions evoke a hydrodynamic
language, our calculations do not require the system to be a proper fluid (and a system of free
particles is certainly not). Indeed, we deal
with a global thermodynamic equilibrium situation which - in the form of the density operator
\eqref{general}, applies to any system, either very weakly or very
strongly interacting. Particularly, our calculation for a free Dirac field means that we seek the thermodynamic equilibrium distribution of a gas of quasi-free fermions where the interaction energy - which would be
necessary to achieve equilibrium dynamically - is neglected, just like for the familiar Bose-Einstein 
or Fermi-Dirac distribution functions. Its relevance for phenomenological situations is further addressed 
in section~\ref{summary}.

In ref.~\cite{ScalarField} we put forward a method to calculate the exact expression of expectation values
with the density operator \eqref{general} (henceforth denoted as thermal expectation values or TEV), based 
on the factorization of the exponentials in the density operator \eqref{general} and their analytical 
continuation to complex thermal vorticity. We introduced a mathematical prescription, dubbed as 
\emph{analytic distillation}, to remove the unphysical, non-analytic terms, which makes it possible to 
obtain the actual physical TEVs by continuing back to real thermal vorticity. In this work, we shall 
extend this method to the fermionic case.

The paper is organized as follows.
In section \ref{sec:basics} we review the spinor formalism for the Dirac field used throughout the paper;
in section~\ref{sec:WigDirac} we review the definition and properties of the Wigner function for the 
Dirac field; in section~\ref{sec:ExactWig} we derive the exact form of the TEVs of quadratic 
combinations of creation and annihilation operators for a particle with arbitrary spin $S$. Then, 
we obtain the exact form of the Wigner function of a free Dirac field and, in section~\ref{sec:currents}, 
starting from the Wigner function, we derive the exact form of the TEVs of the stress-energy tensor, 
the vector current and the axial current. In section~\ref{sec:distillation} we introduce and discuss
the analytic distillation procedure for the fermionic case and in section~\ref{sec:meanvalues} we 
present the results for the two cases of pure acceleration and pure rotation for the massless free 
Dirac field. In section \ref{sec:CKT} we derive the exact form of the distribution function for a 
massless fermions and discuss its relevance for chiral kinetic theory. Finally, in section~\ref{sec:spindensity} 
we derive the exact global equilibrium formula of the spin density matrix and the spin polarization vector.

\subsection*{Notation}
In this paper we use the natural units, with $\hbar=c=K=1$. The Minkowskian metric tensor $g$ is 
${\rm diag}(1,-1,-1,-1)$; for the Levi-Civita symbol we use the convention $\epsilon^{0123}=1$.\\ 
We will use the relativistic notation with repeated indices assumed to be saturated. Operators in Hilbert 
space will be denoted by a wide upper hat, e.g. $\widehat H$, except the Dirac field operator which is
denoted by a $\Psi$. The symbol $\Tr$ will denote the trace over the Hilbert space of quantum field
states, while $\tr$ over a finite dimensional vector space.

\section{Basics of the free Dirac field}
\label{sec:basics}

Before we get into the main topic of this work, it is necessary to introduce the basic formalism of
the Dirac and spin $1/2$ particles theory. We adopt a group theoretical framework, which is very 
convenient for our approach. More details about the formalism can be found in the appendix 
\ref{Spinorial representation}.

It is known that the free Dirac field can be expanded in plane waves:
\begin{equation}
\label{field expansion}
\Psi(x)=\frac{1}{(2\pi)^{3/2}}\sum_{s=-1/2}^{1/2}\int\frac{\di^3\p}{2\varepsilon}
\left[\wa{p}{s} u_s(p)\e^{-ip\cdot x}+\wbd{p}{s} v_s(p)\e^{ip\cdot x}
\right],
 \end{equation}
where $\wa{p}{s}$ and $\wbd{p}{s}$ are the annihilation and creation operators for a particle 
and antiparticle of polarization state $s$ (either a spin component or helicity) and momentum $p$;
$\varepsilon$ is the energy $\sqrt{{\bf p}^2+m^2}$. In the \eqref{field expansion}, the covariant 
anti-commutation rules are implied:
\begin{align*}
     &\{\wa{p}{s},\wad{p'}{t}\}=2\varepsilon \, \delta_{st}\,\delta^3(\mathbf{p}-\mathbf{p'}),\\
     &\{\wb{p}{s},\wbd{p'}{t}\}=2\varepsilon \, \delta_{st}\,\delta^3(\mathbf{p}-\mathbf{p'}).
\end{align*}
The spinors $u_s(p)$ and $v_s(p)$ are column vectors with four components. To make derivations more
compact, it is convenient to introduce $4 \times 2$ matrices $U(p)$ and $V(p)$ for the spinors, as well 
as arranging the creation and destruction operators in column vectors:
\be\label{spinors}
  U(p) = (u_{1/2}(p), u_{-1/2}(p)) \qquad \qquad  
  \wA(p) = \begin{pmatrix} \wa{p}{{1/2}} \\ \wa{p}{{-1/2}} \end{pmatrix} 
\ee
and similarly for $V(p)$ and $\wB^\dagger(p)$. With the notation \eqref{spinors}, the field expansion
can be rewritten as:
\begin{equation}\label{field expansion compact}
\Psi(x)=\frac{1}{(2\pi)^{3/2}} \int\frac{\di^3\p}{2\varepsilon}
\left[ U(p) \wA(p) \, \e^{-\ii p\cdot x} + V (p) \wB^\dagger(p) \, \e^{\ii p\cdot x} \right],
 \end{equation}
with $U(p) \wA(p)$ which is now a multiplication of a matrix $ 4 \times 2$ by a $2 \times 1$.

The spinors $U(p)$ and $V(p)$ are related to the so-called {\em standard Lorentz transformation}, 
henceforth denoted by $[p]$, that transforms a chosen standard four-vector $\mathfrak{p}$ into the
four-momentum $p$.  The spinor form is dictated by the request for the field $\Psi(x)$
to transform according to the  irreducible representation $(0,1/2) \oplus (1/2,0)$ of the
orthochronous Lorentz group ${\rm SO}(1,3)^\uparrow$ \cite{Weinberg:1995mt}.
Therefore, they can be obtained through the application of the representation of the standard Lorentz 
transformation to the spinor $U(\mathfrak{p})$ associated to the standard four-vector $\mathfrak{p}$ 
(see appendix~\ref{Spinorial representation}):
\be\label{spinors2}
 U(p) = \begin{pmatrix} D([p]) & 0 \\ 0 & D([p])^{\dagger-1} \end{pmatrix} U(\mathfrak{p}),
 \qquad \qquad  V(p) = \begin{pmatrix} D([p]) & 0 \\ 0 & D([p])^{\dagger-1} \end{pmatrix} V(\mathfrak{p}),
\ee
where $D( \; )$ stands for the 2-dimensional $(0,1/2)$ representation $D^{(0,1/2)}$ 
of ${\rm SO}(1,3)^\uparrow$. In general, the matrix:
\be\label{slambda1}
  S(\Lambda) = \begin{pmatrix} D(\Lambda) & 0 \\ 0 & D(\Lambda)^{\dagger-1} \end{pmatrix} 
\ee
is the representation of a Lorentz transformation in the $(0,1/2) \oplus (1/2,0)$ full representation
of the Lorentz group including reflections. The generators of ${\rm SO}(1,3)^\uparrow$ in this 
representations are given by:
\begin{equation*}
    \left(\begin{matrix}
    &D^{(0,1/2)}(J^{\mu\nu}) &0\\
    &0 &D^{(1/2,0)}(J^{\mu\nu})
    \end{matrix}\right) =\frac{i}{4}[\gamma^{\mu},\gamma^{\nu}] \equiv \Sigma^{\mu\nu} 
\end{equation*}
so that:
\begin{equation}\label{slambda}
       S(\Lambda) = \exp \left[ -\ii \frac{\phi^{\mu\nu}}{2} \Sigma_{\mu\nu} \right]
\end{equation}
and where the $\gamma$ matrices are those in the so-called Weyl representation:
\begin{align}\label{gammas}
    &\gamma^\mu=\left(\begin{matrix} 0 & \sigma^\mu\\ \bar{\sigma}^\mu &0 \end{matrix}\right)\, , 
    &&\gamma_5=\left(\begin{matrix} I & 0\\ 0 &-I \end{matrix}\right)\, ,
\end{align}
with $\sigma_\mu=(I,\sigma_1,\sigma_2,\sigma_3)$, $\sigma_i$ being the Pauli matrices, 
$\bar\sigma_\mu=(I,-\sigma_1,-\sigma_2,-\sigma_3)$ and $\sigma^\mu = \eta^{\mu\nu} \sigma_\nu$. 
The spinors $U(\mathfrak{p})$, $V(\mathfrak{p})$ in eq.~\eqref{spinors2} read:
\be\label{standarspinors}
   U(\mathfrak{p}) = N \begin{pmatrix} \unt{\mathfrak{p}} \\ \upt{\mathfrak{p}} \end{pmatrix}, \qquad \qquad
   V(\mathfrak{p}) = N \begin{pmatrix} \unt{\mathfrak{p}}C^{-1} \\ \upt{\mathfrak{p}}C \end{pmatrix}.
\ee
In the \eqref{standarspinors}, $N$ is a normalization factor and $C=i\sigma_2$; the lower arc stands
for the hermitian matrix corresponding to a four-vector:
\be\label{varc}
  \unt{X} \equiv X^\mu \sigma_\mu \, ,
\ee
where the upper arc stands for the hermitian matrix corresponding to the reflected vector:
\be\label{varc2}
 \upt{X}=X^\mu \bar\sigma_\mu \, ;
\ee
such definitions are well known in the construction of the ${\rm SL}(2,\mathbb{C})-{\rm SO}(1,3)^\uparrow$ morphism \cite{tung1985group}. The appearance of the matrix $C$ in the $V$ spinor is dictated by the required transformation properties of the $\wA$
and $\wB$ operators under charge conjugation. 

The standard vector $\mathfrak{p}$ and the normalization factor $N$ depend on whether $m \ne 0$ or $m=0$. 
In the massive case, $\mathfrak{p}^\mu=(m,{\bf 0})$, $N=1/\sqrt{m}$ and the spinor $U(\mathfrak{p})$ is then:
$$
  U(\mathfrak{p}) = \sqrt{m} \begin{pmatrix} I \\ I \end{pmatrix}
$$
with $I$ the $2 \times 2$ identity matrix. On the other hand, in the massless case, the standard vector
it is usually chosen to be $\mathfrak{p}^\mu=(\kappa,0,0,\kappa)$, where $\kappa$ is some fixed energy 
value. The normalization factor is $1/\sqrt{2 \kappa}$ and the spinor $U(\mathfrak{p})$ reads:
$$
 U(\mathfrak{p}) = \frac{1}{\sqrt{2\kappa}} \begin{pmatrix} 2\kappa & 0 \\ 0 & 0 \\ 0 & 0 \\ 0 & 2\kappa \end{pmatrix}
 = \sqrt{2\kappa} \begin{pmatrix} 1 & 0 \\ 0 & 0 \\ 0 & 0 \\ 0 & 1 \end{pmatrix}.
$$
Because of the above form, it turns out the column spinor with helicity $+1/2$, that is $u_{+}(p)$ has only the 
upper two components non-vanishing and it is the first column of the matrix $U(p)$. Conversely, the 
spinor $u_{-}(p)$ with helicity $-1/2$ is the second column of the $U(p)$ and has the upper two-component
vanishing.

It should be stressed that the spinors depend, as it is apparent from their definition \eqref{spinors}, on 
both the chosen standard Lorentz transformation $[p]$ and the standard vector $\mathfrak{p}$. Nevertheless, 
also the particle states \cite{MoussaStora} - that is the creation and annihilation operators - depend on the 
particular choice of $[p]$ and $\mathfrak{p}$, so as to make the combination $U(p) \wA (p)$ and $V(p) \wB^\dagger(p)$,
hence the field \eqref{field expansion compact}, independent thereof.

The spinors $U(p)$ and $V(p)$, however constructed, fulfill the Dirac equation in momentum space,
i.e. with $\slashed{p}=\gamma^\mu p_\mu$:
$$
   (\slashed p - m) U(p) = 0, \qquad \qquad      (\slashed p + m ) V(p) = 0,
$$
what can be proved by taking into account the ${\rm SL}(2,\mathbb{C})-{\rm SO}(1,3)^\uparrow$ 
correspondence (see appendix \ref{Spinorial representation} for the proof). Note that, from \eqref{gammas}
and \eqref{varc},\eqref{varc2}:
$$
  \slashed p = \begin{pmatrix} 0 & \unt{p} \\ \upt{p} & 0 \end{pmatrix}.
$$

It is possible to write the spinors in a convenient fashion by picking a particular standard Lorentz 
transformation $[p]$. For the massive case, with $\mathfrak{p}^\mu = (m, {\bf 0})$ and $[p]$ the pure 
boost taking $\mathfrak{p}$ to $p$, one has (see appendix \ref{Spinorial representation}):
$$
   U(p) = \frac{1}{\sqrt{2(\varepsilon+m)}} (\slashed{p} + m) \begin{pmatrix} I \\ I \end{pmatrix}.
$$
While for the massless case, by choosing $\mathfrak{p}^\mu = (\kappa, 0, 0, \kappa)$ and 
$[p]$ the composition of a Lorentz boost along the $z$ axis and a rotation taking $\hat {\bf k}$
into $\hat {\bf p}$ with axis $\hat {\bf k} \times \hat {\bf p}$ (see appendix~\ref{Spinorial representation}):
$$
   U(p) = \frac{1}{\sqrt{2 p \cdot \mathfrak{\bar p}}} \slashed{p} \gamma^0  U(\mathfrak{p}),
$$
where the bar over a four-vector implies the reflection of its space components, that is $\bar X^\mu = 
(X^0, -{\bf X})$. With this compact form, it can be readily checked that the spinors fulfill the following 
relations:
\begin{align}\label{unorma}
  & \bar U(p) U(p) = 2m I     &&U(p) \bar U(p) = (\slashed p + m) \\ \nonumber
  & \bar V(p) V(p) = -2m I    &&V(p) \bar V(p) = (\slashed{p}-m)
\end{align}
in both the massive and the massless case. 

It is also convenient to define a matrix $\mathcal{C} = \ii \gamma^2$ which is very useful to
work out most of the expressions involving the anti-particle spinors $V(p)$ for, from the 
\eqref{standarspinors} and the definitions of the $\gamma$'s in \eqref{gammas}:
$$
      V(p) = \mathcal{C} U(p)^* .
$$      
This matrix is involved in two more useful relations: 
\be\label{properties}
  \gamma^0 \mathcal{C} S(\Lambda)^T \mathcal{C} \gamma^0 = S(\Lambda)^{-1},
  \qquad \qquad \gamma^0 \mathcal{C} \gamma^{\mu T} \mathcal{C} \gamma^0 = - \gamma^\mu .
\ee
%

\section{The covariant Wigner function of the free Dirac field}
\label{sec:WigDirac}

All relevant expectation values in statistical quantum field theory can be obtained from the covariant 
Wigner function. For the free Dirac field, this is defined as~\cite{DeGroot:1980dk}:
\begin{equation}\label{Wigner function definition}
    W_{AB}(x,k)=-\frac{1}{(2\pi)^4}\int\di^4y \, \e^{-ik\cdot y}\,\Tr \left( \wrho :\Psi_A\left({\small x-\frac{y}{2}}\right)\overline{\Psi}_B \left({\small x+\frac{y}{2}}\right):\right) ,
\end{equation}
where the colons imply normal ordering, $\overline{\Psi}$ denotes the Dirac conjugate field $\overline{\Psi}=\Psi^\dagger\gamma^0$, and $A,B$ are the spinorial indices, running from 1 to 4. 
The Wigner function \eqref{Wigner function definition} is a $ 4 \times 4$ matrix fulfilling:
$$
W^\dagger(x,k)=\gamma^0W(x,k)\gamma^0. 
$$
The argument $k$ is a general four-vector and it is not on-shell, even in the free field case.
Indeed, by plugging \eqref{field expansion} into \eqref{Wigner function definition}, one obtains
\footnote{Henceforth, $\langle \widehat X \rangle$ is a shorthand for $\Tr (\wrho \widehat X)$}: 
\begin{equation}\label{Wigner function expanded}
    \begin{split}
W(x,k)=&\frac{1}{(2\pi)^3} \sum_{s,t} \int \frac{\di^3\p}{2\varepsilon}\frac{\di^3\p'}{2\varepsilon'} 
\left\{\e^{\ii x\cdot(p'-p)}\left[\langle\wad{p'}{t} \wa{p}{s}\rangle
    u_s(p)\bar{u}_t(p')\delta^4\left(k-\tfrac{p+p'}{2}\right)+\right.\right. \\
& \left.-\langle \wbd{p'}{s}\wb{p}{t}\rangle
    v_s(p')\bar{v}_t(p)\delta^4\left(k+\tfrac{p+p'}{2}\right)\right]
-\left[\e^{\ii (p+p')\cdot x}\langle \wbd{p'}{t}\wad{p}{s}
    \rangle v_t{(p')}\bar{u}_s(p)+\right.\\
&\left.\left.+\e^{-\ii (p+p')\cdot x}  \langle \wa{p}{t}\wb{p'}{s} \rangle
    u_t(p)\bar{v}_s(p')\right]\delta^4\left(k-\tfrac{p-p'}{2}\right)\right\}. 
\end{split}
\end{equation}
The expansion \eqref{Wigner function expanded} makes it apparent that the Wigner function of a free field 
can be decomposed into particle, antiparticle and space-like terms:
\begin{equation}\label{wigdec}
    W(x,k)=\theta(k^2)\theta(k^0)W_+(x,k)+\theta(k^2)\theta(-k_0)W_-(x,k)+\theta(-k^2)W_S(x,k),
\end{equation}
which can be singled out multiplying the Wigner function $W(x,k)$ by appropriate Heaviside $\theta$ 
functions. 

Using the Dirac equation and integrating by parts, one can show that $W$ is a solution of a partial
differential equation, known as \emph{Wigner equation}:
\begin{equation}\label{Wigner equation}
    \left(\frac{i}{2}\slashed{\partial}+\slashed{k}-m\right)W(x,k)=0.
\end{equation}
It should be stressed that the Wigner equation is independent of the statistical operator, meaning 
that the equation provides no information about the thermodynamic equilibrium value. The Wigner 
equation is, in this respect, only a constraint that the physical $W(x,k)$ must fulfill. Thus far, 
the only known non-trivial solution of this equation is the one associated with global homogeneous 
equilibrium, at constant four-temperature. The form derived in this work, the Wigner function at 
global thermodynamic equilibrium with acceleration and rotation is a new, non-trivial, solution
of the \eqref{Wigner equation}.

The most important feature of the Wigner function is that it allows to express the expectation values of 
local operators as integrals in $k$. For instance, the mean values of the vector current, axial current, 
and canonical stress-energy tensor of the free Dirac field can be written as:
\begin{subequations}
\label{wignerint}
\begin{align}
& j^\mu(x) \equiv \langle:\wj^\mu(x):\rangle=\langle:\overline{\Psi}(x)\gamma^\mu\Psi(x):\rangle=
\tr\left(\gamma^\mu\int\di^4k \,W(x,k)\right),\label{mean current}\\
& j_A^\mu(x) \equiv \langle:\wj\subs{A}^\mu(x):\rangle=\langle:\overline{\Psi}(x)\gamma^\mu\gamma_5\Psi(x):\rangle=
\tr\left(\gamma^\mu\gamma_5\int\di^4k \,W(x,k)\right),\label{mean axial current}\\
& T_C^{\mu\nu}(x) \equiv \langle:\wT^{\mu\nu}_C(x):\rangle=\langle:\frac{i}{2}(\overline{\Psi}(x)
\gamma^\mu\codevnu \Psi(x):\rangle=\tr\left(\gamma^\mu\int\di^4k \,k^\nu W(x,k)\right),
\label{meanset}
\end{align}
\end{subequations}
where $\codevnu={\stackrel{\rightarrow}{\partial^\nu}}-{\stackrel{\leftarrow}{\partial^\nu}}$.

Such expressions are very useful, but since the pseudo-momentum variable $k$ is off-shell, as we have
seen, it is desirable, in several circumstances, to deal with a distribution function with only on-shell
momentum as argument, like in the classical kinetic theory. 
To define it starting from the Wigner function, one could follow the method used in the scalar field case
\cite{ScalarField}, namely recast the mean current \eqref{mean current} as an integral over on-shell momenta
multiplied by a four-vector $p^\mu$. This method can be applied to particles and antiparticles 
separately by taking advantage of the decomposition \eqref{wigdec}; one can then focus on the particle 
contribution only, the antiparticle being easily obtained from it. By using the \eqref{Wigner function expanded}
into the definition \eqref{mean current}, we can then write the particle contribution to the mean current as:
\begin{align}\label{jpart}
    j^\mu_+(x) =& \frac{1}{(2\pi)^3}\sum_{s,t}\int \frac{\di^3\p}{2\varepsilon}\frac{\di^3\p'}{2\varepsilon'}
    \, \e^{\ii (p'-p)\cdot x}\langle \wad{p'}{t} \wa{p}{s}\rangle \, \bar{u}_t(p')\gamma^\mu u_s(p) .
\end{align}
The next step would be to write the above current in terms of a distribution function $f(x,p)$ with on-shell 
momenta. In general, the current in \eqref{jpart} can be written as a momentum integral of a ``phase-space"
vector field depending on both momentum and space-time point:
\be\label{phspacecurrent}
     j^\mu_+(x) = \int \frac{\di^3 \p}{\varepsilon}  {\cal J}^\mu(x,p) .
\ee
Nevertheless, unlike in a classical theory, in general $\mathcal{J}$ is not directed along $p$.
Hence, in order to identify a suitable distribution function $f(x,p)$, one has to decompose the
phase space vector field into the sum of a term parallel to $p^\mu$ and terms orthogonal to it:
$$
 {\cal J}^\mu(x,p) = p^\mu f(x,p) + N^\mu(x,p),    \qquad \qquad N \cdot p = 0 .
$$
For the massive Dirac field, one can obtain such decomposition simply by applying the Gordon identity.
On the other hand, the massless case presents some interesting complications that are dealt with
in section \ref{sec:CKT} in order to compare with some recent achievements in chiral kinetic theory.

\section{Exact Wigner function at global thermodynamic equilibrium}
\label{sec:ExactWig}

In this section, an exact form of the Wigner function~\eqref{Wigner function definition} 
for the general global thermodynamic equilibrium density operator~\eqref{general} will be derived. 
It will be obtained with the same method used in ref.~\cite{ScalarField}: a factorization of the 
density operator and an iterative procedure to calculate the expectation values 
$\langle \wad{p}{s} \wa{p^\prime}{t}\rangle$ with imaginary thermal vorticity. The integrals of 
the Wigner function \eqref{wignerint} are then analytically continued to real thermal vorticity to 
obtain the physical values. For the sake of simplicity, we set $\zeta=0$ in the density operator~\eqref{general}; the extension to a finite $\zeta$ is straightforward. 

\subsection{Expectation values}

As has been mentioned, the method presented in \cite{ScalarField} involves two steps.
The first step is the factorization of the density operator \eqref{general}:
\begin{equation}  \label{factorized rho}
    \wrho=\frac{1}{Z}\exp\left[ -b_\mu \wP^\mu+\frac{\varpi_{\mu\nu}}{2}\wJ^{\mu\nu} \right] = 
    \frac{1}{Z} \exp\left[ {-\widetilde{b}_\mu(\varpi)\wP^\mu} \right]
     \exp\left[ {\frac{\varpi_{\mu\nu}}{2}\wJ^{\mu\nu}} \right],
\end{equation}
where the tilde transform of a vector is defined as:
\begin{equation}\label{tilde transf definition}
    \widetilde{b}^\mu(\varpi)=\sum_{k=0}^\infty\frac{i^k}{(k+1)!}
    (\underbrace{\varpi^\mu_{\ \alpha_1}\varpi^{\alpha_1}_{\ \alpha_2}\dots\varpi^{\alpha_{k-1}}_{\ \ \ \alpha_k}}_{k 
    \text{ times}})b^{\alpha_k}.
\end{equation}
To obtain a finite expression of $\langle \wad{p}{s} \wa{p^\prime}{t}\rangle$ it is necessary
to deal with an imaginary thermal vorticity, setting $\varpi_{\mu\nu}=-i\phi_{\mu\nu}$. Thereby,
one can deal with an actual Lorentz transformation $\widehat{\Lambda}=\exp[-\ii\phi:\widehat J/2]$
and the density operator \eqref{factorized rho} becomes a product of an exponential of the four-momentum 
times the unitary representation of a Lorentz transformation in the Hilbert space:
\begin{equation*}
    \wrho = \frac{1}{Z} \exp\left[ {-\widetilde{b}_\mu(-\ii \phi) \wP^\mu} \right]
     \exp\left[ - \ii {\frac{\phi_{\mu\nu}}{2}\wJ^{\mu\nu}} \right] = 
     \frac{1}{Z} \exp\left[ {-\widetilde{b}_\mu(-\ii \phi) \wP^\mu} \right] {\widehat \Lambda} .
\end{equation*}
This form makes it possible to use Poincar\'e group representation theory to calculate 
$\langle \wad{p}{s} \wa{p^\prime}{t}\rangle$. Let $S$ be the spin of a particle (or helicity, for 
massless particles) and $s$ its polarization state label running from $-S$ to $S$ ($\pm S$ for massless
particles). A creation operator transforms under $\widehat\Lambda$ as:
\be\label{transrule}
    \widehat{\Lambda}\wad{p}{s} \widehat{\Lambda}^{\dagger}=\sum_r D^S(W(\Lambda,p))_{rs}\wad{\Lambda p}{r},
\ee
where $W(\Lambda,p)=[\Lambda p]^{-1}\Lambda[p]$ is a member of the so-called {\em little group}, that is
the subgroup of Lorentz transformations leaving the standard momentum $\mathfrak{p}$ invariant; $D^S$ 
stands for the $(0,S)$-th finite-dimensional representation of the Lorentz group (see appendix 
\ref{Spinorial representation}). 
As we have seen, for massive particles $\mathfrak{p}=(m,\bm{0})$ and the little group is just SO(3); 
in this case, the $W(\Lambda,p)$ matrix is the so-called Wigner rotation. For massless particles, the 
little group is the euclidean group (rotations and translations) in two dimensions, the ISO(2) group;
yet, the translations do not play a physical role and the transformation $D(W(\Lambda,p))$ appearing 
in \eqref{transrule} boils down to a phase: $D(W(\Lambda,p))_{rs}=\e^{-ir\theta(\Lambda,p)}\delta_{rs}$ 
\cite{Weinberg:1995mt}.

The second step to calculate $\langle \wad{p}{s} \wa{p}{t}\rangle$ involves the use of commutation 
or anticommutation relations and the transformation rule \eqref{transrule}:
\begin{equation}\label{aaeq}
\begin{split}
\langle \wad{p}{s} \wa{p'}{t}\rangle=&\frac{1}{Z}\Tr\left(\e^{-\widetilde{b}\cdot \wP}\widehat{\Lambda} \wad{p}{s} 
 \wa{p'}{t}\right)
=\frac{1}{Z}\sum_r D^S(W(\Lambda,p))_{rs}\Tr\left(\e^{-\widetilde{b}\cdot\wP}\wad{\Lambda p}{r}\widehat{\Lambda} 
 \wa{p'}{t}\right)\\
=&\sum_r D^S(W(\Lambda,p))_{rs}\e^{-\widetilde{b}\cdot\Lambda p}\langle \wa{p'}{t} \wad{\Lambda p}{r} \rangle \\
=&(-1)^{2S}\sum_r D^S(W(\Lambda,p))_{rs}\e^{-\widetilde{b}\cdot\Lambda p}\langle \wad{\Lambda p}{r}  \wa{p'}{t} \rangle+\\
&+2\varepsilon \, \e^{-\widetilde{b}\cdot \Lambda p} D^S(W(\Lambda,p))_{ts}\delta^3(\Lambda\bm{p}-\bm{p}'),
\end{split}
\end{equation}
where the factor $(-1)^{2S}$ selects commutation or anticommutation for integer or half-integer spin respectively. A special solution of the equation~\eqref{aaeq} can be 
found by means of the iterative method described in ref.~\cite{ScalarField}. One starts by taking the rightmost term in the \eqref{aaeq} as the leading order approximation of the expectation value:
\begin{equation*}
    \langle \wad{p}{s} \wa{p'}{t}\rangle \sim 2\varepsilon \, \e^{-\widetilde{b}\cdot \Lambda p} 
    D^S(W(\Lambda,p))_{ts}\delta^3(\Lambda\bm{p}-\bm{p}').
\end{equation*}
Then, the leading order solution is fed into the right hand side of the \eqref{aaeq} to get 
an updated expression of the solution, which is the sum of two terms involving delta-functions:
\begin{equation*}
\begin{split}
\langle \wad{p}{s} \wa{p'}{t}\rangle \sim& 2\varepsilon \, (-1)^{2S}\sum_r D^S(W(\Lambda,p))_{rs}
D^S(W(\Lambda,\Lambda p))_{tr}\e^{-\widetilde{b}\cdot\left(\Lambda p+\Lambda^2 p\right)} \delta^3(\Lambda^2\bm{p}-\bm{p}')+ \\
&+2\varepsilon \, \e^{-\widetilde{b}\cdot \Lambda p} D^S(W(\Lambda,p))_{ts}\delta^3(\Lambda\bm{p}-\bm{p}').
\end{split}
\end{equation*}
The product of two transformations of the little group can be written in a more compact form:
\begin{equation*}
    D^S(W(\Lambda,\Lambda p))D^S(W(\Lambda, p))=
    D^S([\Lambda \Lambda p]^{-1}\Lambda[\Lambda p][\Lambda p]^{-1}\Lambda[p])=D^S(W(\Lambda^2,p)),
\end{equation*}
and the updated solution becomes:
\begin{equation*}
\begin{split}
    \langle \wad{p}{s} \wa{p'}{t}\rangle \sim& 2\varepsilon \, (-1)^{2S}D^S(W(\Lambda^2, p))_{ts}
    \e^{-\widetilde{b}\cdot\left(\Lambda p+\Lambda^2 p\right)} \delta^3(\Lambda^2\bm{p}-\bm{p}')+ \\
    &+2\varepsilon \, \e^{-\widetilde{b}\cdot \Lambda p} D^S(W(\Lambda,p))_{ts}\delta^3(\Lambda\bm{p}-\bm{p}').
\end{split}
\end{equation*}
Iterating, one eventually obtains a series of delta functions:
\begin{equation}\label{number operator iterative result}
    \langle \wad{p}{s} \wa{p'}{t}\rangle=2\varepsilon' \, \displaystyle \sum_{n=1}^{\infty}(-1)^{2S(n+1)}
    \delta^3(\Lambda^n\bm{p}-\bm{p}')D^S(W(\Lambda^n,p))_{ts}
    \e^{-\widetilde{b}\cdot\sum_{k=1}^n\Lambda^kp}.
\end{equation}
A similar formula can be obtained for the anti-particle quadratic combination 
$\langle \wbd{p}{s}\wb{p'}{t} \rangle$.
As also mentioned in \cite{ScalarField}, all other expectation values of combinations of creation and annihilation
operators, like e.g. $\wad{p}{s}\wad{p'}{t}$, $\wa{p}{s}\wa{p'}{t}$ and mixed particle-antiparticle combinations, vanish 
because the global equilibrium density operator \eqref{general}, depending on conserved generators of symmetry 
transformations, cannot change the number of particles/antiparticles in a given state. This can be confirmed
by an explicit iterative calculation.

In the trivial case of vanishing thermal vorticity $\varpi=0$, which implies $\Lambda=I$,
it is straightforward to sum the right hand side of the eq.~\eqref{number operator iterative result}:
\begin{equation*}
     \langle \wad{p}{s} \wa{p'}{t}\rangle=2\varepsilon' \, \displaystyle \sum_{n=1}^{\infty}(-1)^{2S(n+1)}
     \delta^3(\bm{ p}-\bm{p}')\delta_{ts}\,\e^{-nb\cdot p}=2\varepsilon \, \delta^3(\bm{p}-\bm{p}')\delta_{ts}
     \,\frac{1}{\e^{b\cdot p}+(-1)^{2S+1}}
\end{equation*}
reproducing the familiar Bose-Einstein and the Fermi-Dirac distributions. This makes it clear that
the series in $n$ corresponds to the typical expansion of the quantum statistics and that the $n > 1$
terms are the corrections are the Boltzmann statistics term, which is the first, with $n=1$. 

As has been mentioned, the solution \eqref{number operator iterative result} becomes the physical mean 
value of the number operator at general global equilibrium only \emph{after} the analytic continuation to
real thermal vorticity. However, this continuation cannot be done at the level of 
\eqref{number operator iterative result}, because of the singular delta functions. Nonetheless, the 
\eqref{number operator iterative result} can be used once momentum integration is carried out.

\subsection{The Wigner function}
\label{subsecc: calculations exact Wigner}

We are now in a position to calculate the Wigner function at global equilibrium, for imaginary 
thermal vorticity, both for massive and massless particles. Feeding the \eqref{number operator iterative result} 
into the \eqref{Wigner function expanded} and taking into account the vanishing of the expectation 
values of some combinations as discussed above, one gets:
\begin{equation}
\begin{split}
 W(x,k) & = \frac{1}{(2\pi)^3}\int\frac{\di^3\p}{2\varepsilon}\sum_{n=1}^\infty (-1)^{n+1}
 \e^{-ix\cdot\left(\Lambda^n p-p\right)}\e^{-\widetilde{b}\cdot\sum_{k=1}^n\Lambda^kp} \\ 
& \times \sum_{rs}\left[ u_r(\Lambda^np)D(W(\Lambda^n,p))_{rs}\bar{u}_{s}(p)
  \delta^4\left( k-\tfrac{\Lambda^np+p}{2} \right) \right. \\ 
& \left. -v_{r}(p)D(W(\Lambda^n,p))_{s r}\bar{v}_s(\Lambda^np)\delta^4\left( k+\tfrac{\Lambda^np+p}{2} \right)\right],
\end{split}
\end{equation}
or, using the compact $4\times 2$ spinor notation \eqref{spinors}:
\begin{equation}
\begin{split}
& W(x,k)=\frac{1}{(2\pi)^3}\int\frac{\di^3\p}{2\varepsilon}\sum_{n=1}^\infty (-1)^{n+1}
\e^{-ix\cdot\left(\Lambda^n p-p\right)}\e^{-\widetilde{b}\cdot\sum_{k=1}^n\Lambda^kp}\times \\ 
&\left[U(\Lambda^np)D(W(\Lambda^n,p))\bar{U}(p)\delta^4\left(k-\tfrac{\Lambda^np+p}{2}\right)
-V(p)D(W(\Lambda^n,p))^T \bar{V}(\Lambda^np)\delta^4\left(k+\tfrac{\Lambda^np+p}{2}\right)\right].
\end{split}
\end{equation}
The above expression can be further worked out by using the transformation rules of the spinors 
under a Lorentz transformation (see appendix \ref{Spinorial representation}):
$$ 
  U(\Lambda p) D(W(\Lambda,p))^{\dagger -1} = S(\Lambda) U(p), 
$$  
where $S(\Lambda)$ is given by the eq. \eqref{slambda}, and taking advantage of the unitarity of the $D(W)$ matrices. By using the eq.~\eqref{unorma} and
the basic representation theory rule $S(\Lambda_1\Lambda_2) = S(\Lambda_1) S(\Lambda_2)$, we thus
have:
$$
  S(\Lambda^n) = S(\Lambda)^n
$$
and:
\begin{equation*}
    U(\Lambda^n p)D(W(\Lambda^n,p))\bar{U}(p)=S(\Lambda)^n U(p) \bar U(p) = S(\Lambda)^n (\slashed p + m).
\end{equation*}
Similarly, for antiparticles, keeping in mind that $V^*=\mathcal{C}U$:
\begin{align*}
 V(p) D(W(\Lambda^n,p))^T \bar{V} (\Lambda^n p) = &
 -\left(\gamma_0\mathcal{C}U(\Lambda^n p)W(\Lambda^n,p)\bar{U}(p)\mathcal{C}\gamma^0\right)^T \\       
 =& -\gamma^0\mathcal{C}(m+\slashed{p}^T) {S(\Lambda)^n}^T \mathcal{C}\gamma^0 =
  -(m-\slashed{p})S(\Lambda)^{-n}
\end{align*}
where we have used the relations \eqref{properties}.
Therefore, the Wigner function can be finally written as:
\begin{equation}\label{eq:WignerSol1}
\begin{split}
    W(x,k)=&\frac{1}{(2\pi)^3}\int \frac{\di^3\p}{2\varepsilon} 
    \sum_{n=1}^{\infty}(-1)^{n+1}\e^{-ix\cdot\left(\Lambda^n p-p\right)} 
    \e^{-\widetilde{b}\cdot\sum_{k=1}^n\Lambda^kp}\times\\ 
&\left[S(\Lambda)^n (m+\slashed{p})\delta^4\left( k-\frac{\Lambda^n p+p}{2}\right)
+(m-\slashed{p})S(\Lambda)^{-n}\delta^4\left(k+\frac{\Lambda^np+p}{2}\right)\right].
\end{split}
\end{equation}
All the relations used to obtain the above expression hold in the massive and in the 
massless case as well. Therefore, the Wigner function for a massless Dirac field is 
simply obtained setting $m=0$ in the eq.~\eqref{eq:WignerSol1}.

It can be shown that the series \eqref{eq:WignerSol1} is a non-trivial solution of the Wigner equation 
\eqref{Wigner equation}. We will confine ourselves to the particle term, the calculations for 
the antiparticle one being very similar. The only dependence on the coordinate in the Wigner 
function is in the exponential, so we can write:
\begin{align*}
    \frac{\ii}{2}\slashed{\partial} \mapsto \frac{1}{2}(\slashed{\Lambda^n p} - \slashed p) 
    = \slashed{\Lambda^n p} - \slashed{k},
\end{align*}
where we have taken advantage of the $\delta$ function. 
Now, since the matrices $\unt{p}$ and $\upt{p}$ transform under Lorentz transformation according to 
\begin{equation*}
    D(\Lambda) \unt{p}D(\Lambda)^\dagger=\unt{\Lambda p}, \qquad \qquad D(\Lambda)^{\dagger-1}
    \upt{p}D(\Lambda)^{-1}=\upt{\Lambda p},
\end{equation*}
and using $\unt{p}\upt{p}=m^2$ (see appendix \ref{Spinorial representation}), we can work out the 
matrix product:
\begin{equation*}
    \begin{split}
        &\slashed{\Lambda^n p}S(\Lambda)^n(m+\slashed{p})=
        \left(\begin{matrix}
        &0 & \unt{\Lambda^n p}\\
        &\upt{\Lambda^n p} &0
        \end{matrix}\right)
        \left(\begin{matrix}
        &\Lambda^{n} &0\\
        &0 &{{\Lambda^{n}}^\dagger}^{-1}
        \end{matrix}\right)
        \left(\begin{matrix}
        &m &\unt{p}\\
        &\upt{p} &m
        \end{matrix}\right)=\\
        &\left(\begin{matrix}
        &0 &\Lambda^n \unt{p}\\
        &{{\Lambda^{n}}^\dagger}^{-1}\upt{p}
        \end{matrix}\right)
        \left(\begin{matrix}
        &m &\unt{p}\\
        &\upt{p} &m
        \end{matrix}\right)=
        \left(\begin{matrix}
        &\Lambda^n m^2 &m\Lambda^n\unt{p}\\
        &m{{\Lambda^n}^\dagger}^{-\dagger} &m^2{{\Lambda^n}^\dagger}^{-1}
        \end{matrix}\right)=mS(\Lambda)^n(m+\slashed{p}).
    \end{split}
\end{equation*}
whence it ensues
\begin{equation*}
    \frac{i}{2}\slashed{\partial}W_+=\left(m-\slashed{k}\right)
    W_+,
\end{equation*}
which proves that \eqref{eq:WignerSol1} solves the \eqref{Wigner equation}. 
The form in the eq.~\eqref{eq:WignerSol1} can be further simplified. Defining the tilde transform
of $\beta$ from the eq.~\eqref{fourtemp} as in \eqref{tilde transf definition}
$$
  \widetilde\beta (\varpi) =\sum_{k=0}^\infty\frac{i^k}{(k+1)!}
    (\underbrace{\varpi^\mu_{\ \alpha_1}\varpi^{\alpha_1}_{\ \alpha_2}\dots
    \varpi^{\alpha_{k-1}}_{\ \ \ \alpha_k}}_{k \text{ times}}) \beta^{\alpha_k},
$$
and taking advantage of the two identities proved in ref.~\cite{ScalarField}, namely:
\begin{equation*}
     \sum_{k=1}^n\Lambda^{-k}\widetilde{b}(\varpi)=n\widetilde{b}(-n\varpi)
\end{equation*}
and
\be\label{tildebetax}
     -\ii x\cdot(\Lambda^np-p)-n\widetilde{b}(-n\varpi)\cdot p=-n\widetilde{\beta}(-n\varpi)\cdot p,
\ee
we can rewrite the \eqref{eq:WignerSol1} as:
\begin{equation}\label{exact Wigner function}
\begin{split}
W(x,k)=&\frac{1}{(2\pi)^3}\int \frac{\di^3\p}{2\varepsilon} \sum_{n=1}^{\infty}(-1)^{n+1}
\e^{-n\widetilde{\beta}_n \cdot p}\times \\
&\left[S(\Lambda)^n (m+\slashed{p})\delta^4\left( k-\frac{\Lambda^n p+p}{2} \right)
+(m-\slashed{p})S(\Lambda)^{-n}\delta^4\left(k+\frac{\Lambda^np+p}{2}\right)\right],
\end{split}
\end{equation}
where, henceforth:
$$
 \widetilde \beta_n \equiv \widetilde{\beta}(-n\varpi) .
$$
This is our final expression for the covariant Wigner function for imaginary thermal vorticity;
for the massless case, the expression is obtained by just taking the $m=0$ limit of the 
\eqref{exact Wigner function}. 
Like the expectation values \eqref{number operator iterative result}, it cannot straightforwardly 
be continued to real thermal vorticity because of the singular delta functions. Nevertheless, 
the \eqref{exact Wigner function} can be integrated in either $\di^4 k$ or $\di^3 \p$ and the 
resulting expressions are fit to be continued analytically.

\section{Currents at global thermodynamic equilibrium}
\label{sec:currents}

The mean currents \eqref{wignerint} are integrals of the Wigner function \eqref{exact Wigner function} and, 
as such, we will see that they give rise to non-singular expressions. Let us start by studying the 
vector current $j^\mu(x)$. Of course, in case of vanishing chemical potential, as implied by $\zeta=0$, 
it ought to be zero and yet it is instructive too see how this comes about. Plugging the 
\eqref{exact Wigner function} into the \eqref{mean current} we obtain:
\begin{equation}
\label{eq:VectorSeries}
  j^\mu(x) = \frac{1}{(2\pi)^3}\int\frac{\di^3\p}{2\varepsilon}\sum_{n=1}^\infty(-1)^{n+1}
  \e^{-n\widetilde{\beta}_n\cdot p}\left[\tr\left(\gamma^\mu S(\Lambda)^n\slashed{p}\right)-
  \tr\left(\gamma^\mu\slashed{p}S(\Lambda)^{-n}\right)\right].
\end{equation}
The first term is the particle contribution, whereas the second term is the antiparticle one.
Now, by using known properties of the trace and of the $\gamma^0\mathcal{C}$ matrix:
\begin{equation}
\label{C oddity}
\begin{split}
\tr(\gamma^\mu\slashed{p}S(\Lambda)^{-n})= &\tr((\gamma^\mu\slashed{p}S(\Lambda)^{-n})^T) =  
 \tr\left({{S(\Lambda)^n}^T}^{-1}\slashed{p}^T {\gamma^{\mu}}^T \right) \\
= & \tr\left(\gamma^0\mathcal{C}S(\Lambda)^n\mathcal{C}\gamma^{0}\gamma^0\mathcal{C}\slashed{p}
 \mathcal{C}\gamma^0\gamma^0\mathcal{C}{\gamma^{\mu}}\mathcal{C}\gamma^0\right) =
  \tr(\gamma^\mu S(\Lambda)^n\slashed{p})
  \end{split}
\end{equation}
which proves that the integrand in \eqref{eq:VectorSeries} vanishes.

Using the commutation rules of the gamma matrices, the particle term of the trace
in~\eqref{eq:VectorSeries} can be written as: 
\begin{equation}\label{eq:traceJ}
    \tr\left(\slashed{p}\gamma^\mu S(\Lambda)^n\right)=p^\mu\tr\left(S(\Lambda)^n\right)+
    2 \ii p_\nu\tr\left(\Sigma^{\mu\nu} S(\Lambda)^n\right),
\end{equation}
therefore, the particle term of the vector current is:
\begin{equation}\label{equilibrium particle current}
 j_+^\mu(x) = \frac{1}{(2\pi)^3}\int\frac{\di^3\p}{2\varepsilon}
    \sum_{n=1}^\infty(-1)^{n+1}\e^{-n\widetilde{\beta}_n\cdot p}
    \left[p^\mu\tr\left( S(\Lambda)^n\right)+2i p_\nu\tr\left(\Sigma^{\mu\nu} S(\Lambda)^n\right)\right],
\end{equation}
where we can identify a term proportional to $p^\mu$ and one perpendicular to it 
(see the discussion at the end of section~\ref{sec:WigDirac} and in section~\ref{sec:CKT}).

The series for the axial current~\eqref{mean axial current} can be derived likewise. In this 
case, being the axial current even under charge conjugation, it may be non vanishing with
$\zeta=0$. Indeed, it can be seen that the particle and antiparticle contributions sum up;
from the eq.~\eqref{exact Wigner function}:
\begin{equation*}
     j^\mu_A(x) = \frac{2}{(2\pi)^3}\int\frac{\di^3\p}{2\varepsilon}
     \sum_{n=1}^\infty(-1)^{n+1}\e^{-n\widetilde{\beta}_n\cdot p}
     \tr\left(\gamma^\mu\gamma_{5}S(\Lambda)^n\slashed{p}\right).
\end{equation*}
Like in \eqref{eq:traceJ} the trace can be split:
\begin{equation*}
\tr\left(\slashed{p}\gamma^\mu\gamma_5S(\Lambda)^n\right)=p^\mu\tr\left(\gamma_5S(\Lambda)^n\right)+2i p_\nu\tr\left(\Sigma^{\mu\nu}\gamma_5S(\Lambda)^n\right).
\end{equation*}
It is convenient, for later use, to introduce a compact notation and define:
\begin{align}
 &A^{\mu\nu}(n)=\tr\left(\gamma^\nu\gamma^\mu S(\Lambda)^n\right)=
 g^{\mu\nu}\tr\left(S(\Lambda)^n\right)+2i \tr\left(\Sigma^{\mu\nu}S(\Lambda)^n\right),
 \label{definition A}\\
 &A^{\mu\nu}_5(n)=\tr\left(\gamma^\nu\gamma^\mu\gamma_5S(\Lambda)^n\right)=
g^{\mu\nu}\tr\left(\gamma_5S(\Lambda)^n\right)+2i \tr\left(\Sigma^{\mu\nu}\gamma_5S(\Lambda)^n\right).
\label{definition A5}
\end{align}
and rewrite the axial current accordingly:
\begin{equation}\label{eq:AxialSeries}
\begin{split}
 j^\mu_A(x) &=\frac{2}{(2\pi)^3}\int\frac{\di^3\p}{2\varepsilon}\sum_{n=1}^\infty(-1)^{n+1}
 \e^{-n\widetilde{\beta}_n\cdot p}\tr\left(\gamma^\mu\gamma_{5}S(\Lambda)^n\slashed{p}\right)\\
&=\frac{2}{(2\pi)^3}\int\frac{\di^3\p}{2\varepsilon}\sum_{n=1}^\infty(-1)^{n+1}
\e^{-n\widetilde{\beta}_n\cdot p}p_\nu A^{\mu\nu}_5(n)\\
&=\frac{2}{(2\pi)^3}\sum_{n=1}^\infty\frac{(-1)^{n+1}}{n}A^{\mu\nu}_5(n)
\left(-\frac{\partial}{\partial \widetilde{\beta}_n^\nu}\right)\int\frac{\di^3 \p}{2\varepsilon}
\e^{-n\widetilde{\beta}_n\cdot p}.
\end{split}
\end{equation}
The uniform convergence of the series of the integrands \cite{ScalarField} for imaginary thermal
vorticity makes it possible to exchange the series with the integration. In the massless case, 
the momentum integral is more easily carried out and one gets:
\begin{equation}\label{eq:MasslessInt}
\int\frac{\di^3\p}{2\varepsilon}\e^{-n\widetilde{\beta}_n\cdot p}=\frac{2\pi}{n^2 \widetilde{\beta}_n 
\cdot \widetilde{\beta}_n},
\end{equation}
so that the \eqref{eq:AxialSeries} becomes:
\begin{equation}\label{eq:AxialMassless}
 j^\mu_A(x) =\frac{1}{\pi^2}\sum_{n=1}^\infty
    \frac{(-1)^{n+1}}{n^3} \frac{\widetilde{\beta}_{n\nu}}{(\widetilde{\beta}_n)^4} A^{\mu\nu}_5(n).
\end{equation}

The quantum statistics series can be obtained for the stress-energy tensor as well. 
By using~\eqref{exact Wigner function} in the definition~\eqref{meanset}, one has:
\begin{equation*}
\begin{split}
 T^{\mu\nu}_C &=\int\di^4 k\, k^\nu\tr\left(\gamma^\mu W(x,k)\right)=\\
&=\sum_{n=1}^\infty\frac{(-1)^{n+1}}{(2\pi)^3}\int\frac{\di^3\p}{2\varepsilon}\e^{-n\tilde{\beta}_n\cdot p}\left[
\left(\frac{\Lambda^np+p}{2}\right)^\nu\tr\left(\gamma^\mu S(\Lambda)^n\slashed{p}\right)+
\left(\frac{\Lambda^np+p}{2}\right)^\nu\tr\left(\gamma^\mu\slashed{p}S(\Lambda)^{-n}\right)\right]\\
&=\sum_{n=1}^\infty\frac{2(-1)^{n+1}}{(2\pi)^3}\int\frac{\di^3\p}{2\varepsilon}
\e^{-n\tilde{\beta}_n\cdot p}\left(\frac{\Lambda^np+p}{2}\right)^\nu \tr\left(\gamma^\mu S(\Lambda)^n\slashed{p}\right),
\end{split}
\end{equation*}
where the eq.~\eqref{C oddity} has been used. Now the momenta can be expressed
as derivatives with respect to $\widetilde{\beta}_n$ and, by using the shorthand \eqref{definition A}: 
\begin{equation*}
  T^{\mu\nu}_C  =\sum_{n=1}^\infty\frac{(-1)^{n+1}}{(2\pi)^3n^2}\left(A^{\mu\alpha}(n)
\frac{\partial}{{\partial \widetilde{\beta}_n}_\nu}\frac{\partial}{\partial \widetilde{\beta}^\alpha_n}+
\left(\Lambda^{n}\right)^\nu_{\ \rho}A^{\mu\alpha}(n)\frac{\partial}{{\partial \widetilde{\beta}_n}_\rho}
\frac{\partial}{\partial \widetilde{\beta}_n^\alpha}\right) \int\frac{\di^3\p}{2\varepsilon}
\e^{-n\widetilde{\beta}_n\cdot p}.
\end{equation*}
In the massless case, the momentum integral~\eqref{eq:MasslessInt} is simple and the above
series can be written as:  
\begin{equation}\label{stress energy tensor formula with betatilde}
  T^{\mu\nu}_C =\sum_{n=1}^\infty\frac{(-1)^{n+1}}{2\pi^2}\frac{1}{n^4\widetilde{\beta}_n^4}
\left[ A^{\mu}_{\ \gamma}(n)\left(4\frac{\widetilde{\beta}_n^\nu {{\widetilde{\beta}}_n}^\gamma}
{\widetilde{\beta}_n^2}-g^{\nu\gamma}\right)+\left(\Lambda^n\right)^\nu_{\ \rho} A^{\mu}_{\ \gamma}(n)
\left(4\frac{\widetilde{\beta}_n^\rho {{\widetilde{\beta}}_n}^\gamma}{\widetilde{\beta}_n^2}-g^{\rho\gamma}\right)\right].
\end{equation}

All of the above quantum statistics series have been derived for imaginary $\varpi$, but they no
longer feature, unlike the Wigner function \eqref{exact Wigner function}, singular terms and they 
can, in principle, be continued analytically to the physical real $\varpi$. It is indeed possible 
to obtain finite expressions by Taylor expanding each term of the series in $\varpi$ at some fixed 
order, continue to real $\varpi$ and sum the series in $n$ at each order in $\varpi$. However, just 
as for the scalar field \cite{ScalarField}, it can be realized that this method does not eventually 
lead to a finite result. Indeed, the full series in $n$ are actually divergent, so the above method
of expanding in $\varpi$ and resumming the partial series in $n$ term by term generates an asymptotic 
expansion in $\varpi$. The ultimate reason for the lack of convergence of the full, quantum
statistic series has been deeply investigated in \cite{ScalarField} and it is twofold: on the one hand, 
for pure rotation, the exponential of the density operator $\widehat H - \omega/T_0 \wJ_z$
is not bounded from below unless a boundary condition on the field is set at some finite radius $r < 1/\omega$; 
on the other hand, for pure acceleration, the iterative method which led to the \eqref{number operator iterative result} and ensuing expressions introduces spurious, non-analytic contributions which must be 
subtracted. In order to obtain finite exact results, before analytically continuing to real $\varpi$, 
it is necessary to extract the analytic part of the functions of $\varpi$ around $\varpi=0$ by means 
of the procedure of analytic distillation introduced in ref.~\cite{ScalarField}. 

\section{Fermionic analytic distillation} 
 \label{sec:distillation}

The building block of all the expressions found so far is the solution \eqref{number operator iterative result} 
of the equation \eqref{aaeq}, which was obtained by iteration. However, as it was discussed in
ref.~\cite{ScalarField}, this solution may not be unique. In general, a solution of the \eqref{aaeq} is
the sum of the general solution of the corresponding homogeneous equation 
$$
 \langle \wad{p}{s} \wa{p'}{t}\rangle= (-1)^{2S}\sum_r D^S(W(\Lambda,p))_{rs}
 \e^{-\widetilde{b}\cdot\Lambda p}\langle \wad{\Lambda p}{r}  \wa{p}{t} \rangle
$$
and a special solution of the \eqref{aaeq}. It can be shown that the homogeneous equation cannot
have analytic solutions in $\phi = \ii \varpi = 0$ \cite{ScalarField}. Nevertheless, it may have non-analytic 
solutions and, specifically, they do occur in the pure acceleration case, while in the pure rotation 
case there is no known non-trivial solution of the homogeneous equation. Hence, unless we deal with the 
pure rotation case, the one analytic solution of the \eqref{aaeq} can be determined by subtracting 
the non-analytic contributions in $\phi=\ii \varpi = 0$. 

The mathematical method to make such a subtraction was introduced in ref \cite{ScalarField} and 
was named {\em analytic distillation}. The definition reads:
\begin{definition} 
{\em Let $f(z)$ be a function on a domain $D$ of the complex plane and $z_0 \in \bar D$ a point 
where the function may not be analytic. Suppose that asymptotic\footnote{We denote asymptotic
equality with the symbol $\sim$.} power series of $f(z)$ in $z-z_0$ 
exist in subsets $D_i \subset D$ such that $\cup_i D_i = D$: 
$$
  f(z) \sim \sum_{n} a^{(i)}_n (z-z_0)^n 
$$
where $n$ can take integer negative values. If the series formed with the common coefficients 
in the various subsets restricted to $n \ge 0$ has a positive radius of convergence, the analytic 
function defined by this power series is called analytic distillate of $f(z)$ in $z_0$ and it 
is denoted by $\dist_{z_0} f(z)$.}
\end{definition}

As it is apparent from the above definition, analytic distillation requires the existence of an 
asymptotic power series of the function at the point of interest. However, all series in 
section~\ref{sec:currents} are neither convergent power series nor asymptotic power series, 
rather formal series of trigonometric and hyperbolic functions of $n$-dependent argument (see 
section~\ref{sec:meanvalues}). In the scalar field case \cite{ScalarField}, where one deals with 
series of the form:
\begin{equation*}
 g(x)=\sum_n f(nx)
\end{equation*}
a theorem due to Zagier \cite{zeidler2007quantum,ZagierAppendix} allows to obtain an asymptotic 
power series of the function $g(x)$:
\begin{equation}
\label{eq:scalarAsympt}
g(x) \sim \frac{I_f}{x} + \sum_{n=0}^\infty a_n \zeta(-n) x^n ,\quad
 I_f = \int_0^\infty \di x \; f(x),    
\end{equation}
once the asymptotic power series of $f$ is known:
$$
  f(x) \sim \sum_{n=0}^\infty a_n x^n .
$$ 
However, as it is apparent in section~\ref{sec:currents}, the fermionic series are of the form:
\begin{equation*}
    g(x)=\sum_n(-1)^{n+1}f(nx),
\end{equation*}
and Zagier's theorem does not apply. To handle asymptotic expansions of such alternating series, it
is thus necessary to extend Zagier's theorem and the asymptotic formula. For this purpose, the appropriate 
tool is the generalized Mellin transform, as envisaged in ref.~\cite{zeidler2007quantum}.

Given a function $\varphi(x)$, its Mellin transform is defined as:
\begin{equation}\label{Mellin transform definition}
\{\mathcal{M}\varphi\}(s)=\int_0^\infty\di t \, \varphi(t)\, t^{s-1},
\end{equation}
whence:
$$
\{\mathcal{M}\varphi(\lambda x)\}(s)=\lambda^{-s}\{\mathcal{M}\varphi\}(s) \qquad \qquad \lambda \in \mathbb{R}_{>0} \, .
$$
The Mellin transform defines an holomorphic function of $s$ if $\varphi(x)$ decays rapidly both at 
zero and infinity. If these requirements are not fulfilled, the Mellin transform defines an holomorphic 
function in a smaller region of the complex plane. For instance, if $\varphi(t)\rightarrow t^{-A}$ as 
$t\rightarrow 0$ and $\varphi(t)\rightarrow t^{-B}$ as $t\rightarrow \infty$, the Mellin transform is 
holomorphic in the strip $A<s<B$. In such cases, we can make a meromorphic continuation of the Mellin 
transform over a larger domain. For instance, assume $\varphi$ to be a function of rapid decay at infinity
but to have the following asymptotic expansion in 0:
$$
    \varphi(x)\sim\sum_{n=0} a_n x^{n}.
$$
In this case the Mellin transform defined in \eqref{Mellin transform definition} exists only in the half-plane 
${\rm Re} \, s > 0$. However we can analytically continue the transformed function by defining for some 
$N>0$:
$$
 \{\mathcal{M}\varphi\}(s)=\int_0^1\di t \,\left(\varphi(t)-\sum_{n=0}^{N-1} a_n t^{n}\right)t^{s-1}
 +\sum_{n=0}^{N-1} \frac{a_n}{n+s}+\int_1^\infty\di t\, \varphi(t)t^{s-1}.
$$
This new definition extends Mellin transforms over a larger region of the complex plane ${\rm Re}\,s > - N$, 
and shows that such a continuation is a meromorphic function with simple poles in $s=-n$ with residue $a_n$. 
Conversely, if the generalized Mellin transformation 
$\{\mathcal{M}\varphi\}$ is a meromorphic function with simple poles in $s=-n$, then the coefficients of 
the asymptotic power series of $\varphi$ are the residues of $\{\mathcal{M}\varphi\}$~\cite{ZagierAppendix}. These are the 
basic features we need for the purpose of this work, for further information on the generalized Mellin 
transformation we refer the reader to the refs.~\cite{zeidler2007quantum} and \cite{ZagierAppendix}.

We are now in the position to prove the theorem:
\begin{theorem}\label{th1}
{Let $F(z)$ be a $C^\infty$ complex valued function in a domain of the complex plane and suppose
that $F$ has the following asymptotic power series in $z=0$ 
$$
  F(z) \sim \sum_{k=-M}^\infty A_k z^k
$$
with $M$ a positive integer and let $F$ be $o(1/|z|)$ when $|z| \to +\infty$ in the real axis.
Let $G(z)$ be the function defined by the series:
\begin{equation}\label{G}
   G(z) = \sum_{n=1}^\infty (-1)^{n+1} F(nz).
\end{equation}
The asymptotic power series of $G(z)$ for $|z| \to 0^+$, is given by:
\be\label{eq:AsymptG}
   G(z) \sim \sum_{n=-M}^\infty A_n \eta(-n) z^n ,
\ee
where $\eta$ is the Dirichlet function:
$$
    \eta(s)=\sum_{k=1}^{\infty}\frac{(-1)^{k+1}}{k^s}=(1-2^{1-s})\zeta(s).
$$}
\end{theorem}

To prove the theorem, we first consider a complex valued function with real argument. 
Consider then $f$ on the positive real axis with the following asymptotic power series in $x=0$:
\begin{equation*}
  f(x) \sim \sum_{k=-M}^\infty a_k x^k
\end{equation*}
and let us define the function $g$ as
\begin{equation*}
  g(x) = \sum_{n=1}^\infty (-1)^{n+1} f(nx).
\end{equation*}
It can be shown that $g$ has the following asymptotic power series about $0^+$:
$$
   g(x) \sim \sum_{n=-M}^\infty a_n \eta(-n) x^n.
$$ 
To prove it, let us first remove the negative-powers terms of the asymptotic series of $f$ by 
defining two new functions $\tilde{f}$ and $\tilde{g}$:
\begin{equation*}
\tilde{f}(x) = f(x) - \sum_{k=-M}^{-1} a_k x^k,\quad
\tilde{g}(x) = \sum_{n=1}^\infty (-1)^{n+1} \tilde{f}(nx).
\end{equation*}
It readily follows that
\begin{equation*}
g(x) = \tilde{g}(x) + \sum_{n=1}^\infty\sum_{k=-M}^{-1} (-1)^{n+1} a_k (n x)^k.
\end{equation*}
The finite and infinite sum can be exchanged, yielding:
\begin{equation}
\label{eq:gnegative}
g(x) = \tilde{g}(x) + \sum_{k=-M}^{-1}a_k\eta (-k)x^{k}.
\end{equation} 
We now move to the asymptotic expansion of $\tilde{g}$. From the properties of the generalized 
Mellin transform, $\{\mathcal{M}\tilde{f}\}$ is a meromorphic function having poles in 0 and 
in all the negative integers, with residues $a_n$. Let us now consider the Mellin transform of 
$\tilde{g}$:
\begin{equation*}
    \{\mathcal{M}\tilde{g}\}(s)=\sum_{n=1}^\infty(-1)^{n+1}\{\mathcal{M}\tilde{f}(nx)\}=
    \sum_{n=1}^\infty (-1)^{n+1}n^{-s}\{\mathcal{M}\tilde{f}\}(s)=\eta(s)\{\mathcal{M}\tilde{f}\}(s).
\end{equation*}
Since the $\eta$ function is holomorphic in the complex plane, then $\{\mathcal{M}\tilde{g}\}(s)$ 
is a meromorphic function having simple poles in zero and in all negative integers, just as 
$\{\mathcal{M}\tilde{f}\}$, with residues $\eta(-n)a_n$. Therefore, according to the theorem
proved in ref.~\cite{ZagierAppendix}, $\tilde{g}(t)$ must have the following power asymptotic 
expansion about zero:
\begin{equation*}
    \tilde{g}(x)\sim\sum_{n=0}^\infty a_n\eta(-n)x^n.
\end{equation*} 
Plugging the asymptotic expansion of $\tilde{g}$ in \eqref{eq:gnegative} we finally obtain the asymptotic 
expansion of $g$:
$$
g(x)\sim\sum_{n=-M}^\infty a_n\eta(-n)x^n,
$$
which proves the theorem for a function of real argument. This result can be extended to a 
function of complex variable $F(z)$ with an asymptotic power series about $z=0$:
$$
  F(z) \sim \sum_{n=-M}^\infty A_n z^n.
$$
Defining:
$$
  x = |z|  \implies z = x \, \e^{\ii \varphi} \, ,
$$
and a new complex-valued function $f$ of the real variable $x$ with fixed $\varphi$:
$$
  f(x) \equiv F(x \, \e^{\ii \varphi}).
$$
Now, $f$ has an asymptotic expansion about $x=0$ given by:
$$
f(x) \sim \sum_{n=-M}^\infty A_n \e^{\ii n \varphi} x^n \equiv \sum_{n=-M}^\infty a_n x^n  .
$$
From the previous proof, it follows:
\begin{equation*}
\begin{split}
 G(z) \equiv& \sum_{n=1} (-1)^{n+1} F(nz) = \sum_{n=1} (-1)^{n+1} F(nx \, \e^{\ii \varphi})
 = \sum_{n=1} (-1)^{n+1}  f(nx)\\
 &\sim \sum_{n=-M}^\infty a_n \eta(-n) x^n
 = \sum_{n=-M}^\infty A_n \eta(-n) z^n  ,
\end{split}
\end{equation*}
which proves the theorem in its complete form.\hfill $\square$

We can conclude that, if the asymptotic expansion~\eqref{eq:AsymptG} is convergent, the analytic 
distillate of $G$ is:
$$
  \dist_0 G(z) = \dist_0 \sum_{n=1}^\infty(-1)^{n+1} F(nz) = \sum_{n=0}^\infty A_n \eta(-n) z^n .
$$

The basic idea of the method established by this theorem is to expand in a power series of
$z$ the function $F$ in the series \eqref{G}, to then force the exchange of the two series, and to 
finally replace the divergent expressions with the analytic continuations of the $\eta$ Dirichlet function. 
A remarkable difference with the non-alternating case \eqref{eq:scalarAsympt}, is the absence
of the term proportional to $x^{-1}$; this is owing to the non-singular behaviour of the $\eta$ 
function in $s=1$, at variance with the $\zeta$. This has an interesting consequence, namely that the
full asymptotic series of the function $G(z)$ in \eqref{G} can be obtained in practice by expanding
$f(z)$ in power series (including negative powers) around $z=0$, exchange the series and inserting
the $\eta$ function; in fact, this procedure misses the integral term in the scalar case \cite{ZagierAppendix}.
It is worth mentioning that the $\eta$ function vanishes for all negative even numbers, so one can 
often end up with a finite sum in the distillate.

\section{Exact mean values of currents at global thermodynamic equilibrium}
\label{sec:meanvalues}

We are now in a position to analytically continue the series obtained in section~\ref{sec:currents}
to real thermal vorticity and to obtain finite results by applying the analytic distillation. We
begin with the pure acceleration case and continue with pure rotation and, finally, with a more 
general case.

\subsection{Acceleration}
\label{subsec:acc}

As extensively studied elsewhere \cite{Becattini:2014yxa,Becattini:2017ljh,Buzzegoli:2017cqy,
Buzzegoli:2018wpy,Becattini:2019poj,Prokhorov:2019yft,Prokhorov:2019cik}, 
thermoydnamic equilibrium with constant proper acceleration (along the $z$-axis) is characterized 
by the following thermal vorticity and four-temperature field:
\be\label{eq:thvortAcc}
 \varpi_{\mu\nu}=\frac{a}{T_0}(g_{3\mu}g_{0\nu}-g_{0\mu}g_{3\nu}), 
 \qquad \qquad \beta^\mu=\frac{1}{T_0}\left(1 + a z,0,0, a t\right).
\ee
In this case one can calculate the mean values of scalar operators at $t=z=0$ and obtain them 
in a general space-time point by means of the simple substitution $\beta(0)\rightarrow \beta(x)$
\cite{ScalarField}. For imaginary thermal vorticity, setting $\phi = \ii a/T_0$ we have:
\be\label{eq:btildeAcc}
    n\widetilde{\beta}^\mu_n(t=0,z=0) =\left(\frac{\sinh(n\phi)}{T_0\phi},0,0,\frac{1-\cosh(n\phi)}{T_0\phi}\right),
\ee
hence: 
\begin{equation*}
\beta(0)^2=\frac{1}{T_0^2}, \quad n^2\widetilde{\beta}_n^2=\beta(0)^2\frac{4\sinh^2\left(\frac{n\phi}{2}\right)}{\phi^2}.
\end{equation*}
Using the above expressions, the form of a Lorentz transformation, $S(\Lambda)$ from \eqref{slambda} and the eq.~\eqref{definition A} \eqref{definition A5}, 
we obtain the following matrices:
{\small
\begin{equation}\label{matrices acceleration}
    \begin{split}
    &\left(\Lambda^n\right)^{\mu}_{\ \nu}\!=\!\!\left(\begin{matrix}
    \cosh n\phi &0 &&0 &&&\sinh{n\phi}\\
    0 &1 &&0 &&&0\\
    0 &0 &&1 &&&0\\
    \sinh{n\phi} &0 &&0 &&&\cosh{n\phi}
    \end{matrix}\right)\!, 
    \qquad \qquad S(\Lambda)^n=\left(\begin{matrix}
    &\e^{n\phi/2} &0 & 0 & 0\\
    &0 &\e^{-n\phi/2} &0 & 0\\
    &0 &0 &\e^{-n\phi/2} &0\\
    &0 &0 &0 &\e^{n\phi/2}\\
    \end{matrix}\right),
    \\ &A^{\mu\nu}\!=\!\!\left(\begin{matrix}
    4\cosh\frac{n\phi}{2} &0 &0 &-4\sinh\frac{n\phi}{2}\\
    0 &-4\cosh\frac{n\phi}{2} &0 &0\\
    0 &0 &-4\cosh\frac{n\phi}{2} &0\\
    4\sinh\frac{n\phi}{2} &0 &0 &-4\cosh\frac{n\phi}{2}
    \end{matrix}\right)\!, 
    \quad A^{\mu\nu}_5=\left(\begin{matrix}
    &0 &0 &0 &0\\
    &0 &0 &4i\sinh\frac{n \phi}{2} &0\\
    &0 &-4i\sinh\frac{n \phi}{2} &0 &0\\
    &0 &0 &0 &0
    \end{matrix}\right).
    \end{split}
\end{equation}
}
which, once entered into the eq.~\eqref{stress energy tensor formula with betatilde}, provide us with 
the stress-energy tensor of a free massless Dirac field in $x=0$ with imaginary acceleration $a=-\ii \phi T_0$. Henceforth, the series in $\phi$ for mean values will be denoted with an additional
subscript $I$, for instance:
\begin{equation}
\label{eq:T00Series}
T^{00}_C(0)_I = \frac{3}{8\pi^2}\frac{1}{\beta^4}
    \sum_{n=1}^{\infty}(-1)^{n+1}\phi^4\frac{\sinh{n\phi}}{\sinh^5\left(\frac{n\phi}{2}\right)}
\equiv \frac{3}{8\pi^2}\frac{1}{\beta^4} S_{F,4}(\phi).
\end{equation}
The series $S_{F,4}(\phi)$ uniformly converges when $\phi$ has a non-vanishing real part,
whereas in the physical case of imaginary $\phi$ the  series is badly divergent because of the sine
function in the denominator, thus analytic distillation is required. 
The analytic distillation of a general class of series of the type in eq.~\eqref{eq:T00Series} 
are accurately studied in the appendix~\ref{sec:AccSeries}; here we sketch the specific procedure 
for the series in eq.~\eqref{eq:T00Series}.
The asymptotic power series of the function $T^{00}_C(0)_I $ can be obtained by using theorem~\ref{th1}.
The first step is to derive an asymptotic power series expansion about zero of the function
to be summed, what can be obtained by writing the Laurent series of the ratio
\begin{equation*}
 \frac{\sinh\phi}{\sinh^5\frac{\phi}{2}}=\frac{32}{\phi^4}-\frac{4}{3\phi^2}-\frac{17}{180}
 +\sum_{n=1}^\infty a_n\phi^{2n},
\end{equation*}
which holds true for any complex $\phi$. Since all the positive powers are even, the vanishing of the 
$\eta$-function in the asymptotic formula of theorem \ref{th1} yields a polynomial:
\begin{equation*}
S_{F,4}(\phi)= \phi^4 \sum_{n=1}^{\infty}(-1)^{n+1}
    \frac{\sinh{n\phi}}{\sinh^5\left(\frac{n\phi}{2}\right)}
\sim \phi^4\left( \frac{32\eta(4)}{\phi^4}-\frac{4\eta(2)}{3\phi^2}-\frac{17\eta(0)}{180}\right).
\end{equation*}
Now that we have a converging asymptotic expansion, the analytic distillate of the series is
\begin{equation*}
    \dist_{\phi=0} T^{00}_C(0)_I =\dist_{\phi=0}\left(\frac{3}{8\pi^2}\frac{1}{\beta^4} S_{F,4}(\phi)\right)=\frac{7\pi^2}{60\beta^4}-\frac{\phi^2}{24\beta^4}-\frac{17\phi^4}{960\pi^2\beta^4}.
\end{equation*}
The same method can be used for the other components and one eventually obtains:
\begin{align*}
&\dist_{\phi=0}T^{\text{off diag}}_C=0,\\
&\dist_{\phi=0}T^{11;22;33}_C=\dist_{\phi=0}\left(\frac{1}{8\pi^2}\frac{1}{\beta^4}
S_{F,4}(\phi)\right)=\frac{7\pi^2}{180\beta^4}-\frac{\phi^2}{72\beta^4}-\frac{17\phi^4}{2880\pi^2\beta^4}
\end{align*}
The analytic distillation yields simple polynomials and the analytic continuation to real acceleration 
is readily done by replacing $\phi\rightarrow i a/T_0$. 
    
From these components, we can find the thermodynamic coefficients appearing in the general decomposition 
of the energy-momentum tensor at equilibrium with acceleration \cite{Becattini:2019poj,ScalarField}:
\begin{equation*}
    T^{\mu\nu}_C=\rho u^\mu u^\nu -p(g^{\mu\nu}-u^\mu u^\nu)+\mathcal{A}\alpha^\mu\alpha^\nu,
\end{equation*}
where $u^\mu=\beta^\mu/\sqrt{\beta^2}$ and $\alpha^\mu = \varpi^{\mu\nu} u_\nu$ like in eq.~\eqref{alphaw}. 
After the continuation to the physical case, we find:
\begin{equation}\label{eq:SETCoeffAcc}
\begin{split}
\rho =& \frac{7\pi^2}{60\beta^4}-\frac{\alpha^2}{24\beta^4}-\frac{17\alpha^4}{960\pi^2\beta^4},\\
p =& \frac{7\pi^2}{180\beta^4}-\frac{\alpha^2}{72\beta^4}-\frac{17\alpha^4}{2880\pi^2\beta^4},\\
\mathcal{A}= &0,
\end{split}
\end{equation}
where $\alpha^2=\alpha^\mu\alpha_\mu=-a^2/T_0^2$. According to the remark below the equation
\eqref{eq:thvortAcc}, making $\beta^2$ coordinate-dependent extend the validity of the above
expressions to all space-time.

The results~\eqref{eq:SETCoeffAcc} are in agreement with the perturbative expansion to second order 
in acceleration of ref.~\cite{Buzzegoli:2017cqy,Buzzegoli:2018wpy} and to fourth order of ref.~\cite{Prokhorov:2019yft,Prokhorov:2019cik,Prokhorov2020}. A further confirmation that 
these are the actual exact result is their vanishing at the Unruh temperature $T_U = a/2\pi$
\cite{Becattini:2017ljh,Prokhorov:2019yft}. We note that these formulae upgrade those found in 
ref.~\cite{Florkowski:2018myy} which were based on an approximated form of the Wigner function 
at equilibrium. Therein, a puzzling result was found, that is the vanishing of the energy density
at $T_0=a/\pi$ instead of the Unruh temperature $a/2\pi$. With the exact form of the Wigner 
function found, the agreement with Unruh temperature is fully restored. This feature extends to a 
large class of local operators, see appendix~\ref{sec:AccSeries}. 
Since their mean values are obtained through integration of the Wigner function, we argue that 
the vanishing at the Unruh temperature must extend to any integral of the Wigner 
function~\eqref{exact Wigner function}, after analytic distillation and continuation.
    
\subsection{Rotation}
\label{sec:Rot}

We now turn to the global thermodynamic equilibrium with pure rotation, characterized by:
\begin{align}\label{vorticity rotation}
  \varpi_{\mu\nu}=\frac{\omega}{T_0}(g_{\mu1}g_{\nu2}-g_{\mu2}g_{\nu1}), 
  &&\beta^{\mu}=\frac{1}{T_0}(1,-\omega y,\omega x,0),
\end{align}
 where $\omega$ is the constant angular velocity.
Thus, the density operator of rotational thermodynamic equilibrium
reads:
\be\label{rotdo}
 \wrho = \frac{1}{Z} \exp \left[ -\frac{\widehat H}{T_0} + \frac{1}{T_0} \omegav \cdot \widehat{\bf J} \right].
\ee
The tilde-transformed four-temperature vector for imaginary angular velocity (i.e. $\omega/T_0=-\ii \phi$) is:
$$
 n\widetilde{\beta}(in\phi)=\left(\frac{n}{T_0},ix(\cos(n\phi)-1)+iy\sin(n\phi),iy(\cos(n\phi)-1)-ix\sin(n\phi),0\right),
$$
and its squared magnitude is:
\begin{equation}\label{betarot sqr}
    n^2\widetilde{\beta}_n^2=\frac{n^2}{T_0^2}+4r^2\sin^2\left(\frac{n\phi}{2}\right),
\end{equation}
where $r^2=x^2+y^2$. From \eqref{vorticity rotation}, \eqref{slambda} and \eqref{definition A}, \eqref{definition A5} we obtain:
{\small 
\begin{equation}
\label{eq:RotMatrices}
\begin{split}
&(\Lambda^n)^\mu_{\ \nu}\!=\!\!\left(\begin{matrix}
    1 &0 &0 & 0\\
    0 &\cos n\phi &-\sin n\phi &0\\
    0 &\sin n\phi &\cos n\phi &0\\
    0 &0 &0 &1
\end{matrix}\right), \qquad \qquad
S(\Lambda)^n=\left(\begin{matrix}
    &\e^{-in\phi/2} &0 &0 &0\\
    &0 &\e^{in\phi/2} &0 &0\\
    &0 &0 &\e^{-in\phi/2} &0\\
    &0 &0 &0 &\e^{in\phi/2}
\end{matrix}\right),
\\
&A^{\mu\nu}\!=\!\!\left(\begin{matrix}
    4\cos(\tfrac{n\phi}{2}) &0 &0 &0\\
    0 &-4\cos(\tfrac{n\phi}{2}) &4\sin(\tfrac{n\phi}{2}) &0\\
    0 &-4\sin(\tfrac{n\phi}{2}) &-4\cos(\tfrac{n\phi}{2}) &0\\
    0 & 0 & 0 & -4\cos(\tfrac{n\phi}{2}) 
\end{matrix}\right),\qquad  
A_5^{\mu\nu}=\left(\begin{matrix}
    &0 &0 &0 & 4 i\sin\frac{n\phi}{2}\\
    &0 &0 &0 &0\\
    &0 &0 &0 &0\\
    &-4i\sin\frac{n\phi}{2} &0 &0 &0
\end{matrix}\right).
\end{split}
\end{equation}
}

As an example, we calculate the stress-energy tensor. For comparison with known results in the 
literature, we calculate the symmetrized Belinfante stress-energy tensor, which, for the Dirac 
field, has a simple relation with the previously quoted canonical one:
\begin{equation}\label{belinfante}
    T^{\mu\nu}_B(x)=\frac{1}{2}\left(T^{\mu\nu}_C(x)+T^{\nu\mu}_C(x)\right),
\end{equation} 
and it is thus easy to derive it from \eqref{stress energy tensor formula with betatilde}.
For instance, plugging the~\eqref{betarot sqr} and~\eqref{eq:RotMatrices}
in the $00$-component of~\eqref{stress energy tensor formula with betatilde} we obtain:
\begin{equation}\label{belinfante 00 example}
    T_B^{00}(x)_I =\frac{1}{2\pi^2}\sum_{n=1}^{\infty}\frac{(-1)^{n+1}8T_0^4
    \cos\left(\frac{n\phi}{2}\right)[3n^2+2r^2T_0^2(\cos n\phi-1)]}
    {\left(n^2+4r^2T_0^2\sin^2\left(\frac{n\phi}{2}\right)\right)^3}.
\end{equation}
The series is convergent for real $\phi$, but it becomes divergent as soon as $\phi$ gets
a small imaginary part, just like in the scalar field case \cite{ScalarField} (the 
denominator features infinitely many zeroes, and the series becomes densely divergent in the 
complex plane). Indeed, without boundary condition of the field at a finite radius $r$
such that $\omega r <1$ \cite{ScalarField}, the operator $\widehat H - \omega \wJ_z$ in the 
eq.~\eqref{rotdo} is not bounded from below and gives rise to unphysical divergences. Thus, 
while the analytic distillation procedure leads to finite results, the physical meaning of
rotation without boundary conditions remains limited~\cite{Ambrus:2015lfr}.

Moreover, in the distillation of the function in 
eq.~\eqref{belinfante 00 example}, and for all the series likewise obtained from the Wigner 
function \eqref{exact Wigner function} at thermodynamic equilibrium with rotation, there is an 
additional difficulty. Indeed, as it appears from \eqref{belinfante 00 example}, the series 
does not depend on the product $n\phi$ alone and thus we cannot straightforwardly apply the
theorem~\ref{th1}.
To overcome this problem, we use a method that we proposed in ref.~\cite{ScalarField}. The first
step is to introduce an auxiliary real parameter $B$, and replacing $\phi$ with $B$ any time 
the former is not accompanied by a factor $n$. For the \eqref{belinfante 00 example}:
\begin{equation*}
    T_B^{00}(x)_I =\frac{1}{2\pi^2}\sum_{n=1}^{\infty}\lim_{B\rightarrow\phi}
    \frac{(-1)^{n+1}8B^4T_0^4\cos\left(\frac{n\phi}{2}\right)[3n^2\phi^2+2r^2B^2T_0^2
    (\cos n\phi-1)]}{\left(n^2\phi^2+4r^2B^2T_0^2\sin^2\left(\frac{n\phi}{2}\right)\right)^3}.
\end{equation*}
Since the series is a uniformly convergent series of continuous functions of $\phi$ and $B$ 
for real $B$ and $\phi$, one can exchange the limit with the sum and obtain a series which 
is now suitable for the application of the theorem~\ref{th1}: 
\begin{equation*}
    T_B^{00}(x)_I =\lim_{B\rightarrow\phi}\frac{1}{2\pi^2}\sum_{n=1}^{\infty}
    \frac{(-1)^{n+1}8B^4T_0^4\cos\left(\frac{n\phi}{2}\right)[3n^2\phi^2+2r^2B^2T_0^2(\cos n\phi-1)]}{\left(n^2\phi^2+4r^2B^2T_0^2\sin^2\left(\frac{n\phi}{2}\right)\right)^3}.
\end{equation*}
The asymptotic power series obtained with theorem~\ref{th1} is indeed a finite polynomial in $\phi$:
\begin{equation}\label{asy T00}
    \begin{split}
        T_B^{00}(x)_I \sim \lim_{B\rightarrow \phi} \, \frac{7 \pi ^2 B^4 T_0^4}{60 \phi ^4 \left(1+B^2 r^2 T_0^2\right)^3}-\frac{7 \pi ^2 B^6 r^2 T_0^6}{180 \phi ^4 \left(1+B^2 r^2 T_0^2\right)^3}+\\
        \frac{7 B^6 r^2 T_0^6}{36 \phi ^2 \left(1+B^2 r^2 T_0^2\right)^4}
        -\frac{B^8 r^4 T_0^8}{72 \phi ^2 \left(1+B^2 r^2 T_0^2\right)^4}
       -\frac{B^4 T_0^4}{8
       \phi ^2 \left(1+B^2 r^2 T_0^2\right)^4}+
       \frac{17
       B^{10} r^6 T_0^{10}}{2880 \pi ^2 \left(1+B^2 r^2 T_0^2\right)^5}\\+
       \frac{247 B^8 r^4 T_0^8}{2880 \pi ^2
       \left(1+B^2 r^2 T_0^2\right)^5}
       -\frac{137 B^6 r^2 T_0^6}{576 \pi ^2 \left(1+B^2 r^2 T_0^2\right)^5}+\frac{B^4
       T_0^4}{64 \pi ^2 \left(1+B^2 r^2 T_0^2\right)^5}.
   \end{split}
\end{equation}
Now, one can take the limit $B \to \phi$, thereby obtaining a series which implicitly provides 
the asymptotic power series of the original function \eqref{belinfante 00 example} about $\phi=0$, 
taking into account the analyticity of the various expressions in $B$ in \eqref{asy T00} \cite{ScalarField}.
We can then apply analytic distillation and continue the result to the physical angular velocity 
$\phi \to \ii \omega/T_0$. The final expression is:
\begin{equation*}
\begin{split}
T^{00}_B(x)=&\frac{7}{180} \pi ^2 \left(4 \gamma^2-1\right) \gamma ^4 T_0^4+\frac{1}{72} \left(24 \gamma ^4-16 \gamma ^2+1\right) \gamma ^4 T_0^2 \omega ^2+\\
&+\frac{\left(960 \gamma ^6-1128 \gamma ^4+196 \gamma ^2+17\right) \gamma ^4 \omega ^4}{2880 \pi ^2},
\end{split}
\end{equation*}
where we defined:
\begin{equation*}
    \gamma=\frac{1}{\sqrt{1-r^2\omega^2}}.
\end{equation*}
This procedure can be carried out for all the components of the Belinfante stress-energy
tensor (see appendix \ref{series rotation}); here we just quote the final results.

For this purpose, it is convenient to introduce a tetrad of orthogonal, non-normalized vectors:
\begin{align}\label{tetrad}
    &u^\mu=\frac{\beta^\mu}{\sqrt{\beta^2}}, &&\alpha^\mu=\varpi^{\mu\nu}u_\nu, &&&w^{\mu}=-\frac{1}{2}\epsilon^{\mu\nu\rho\sigma}\varpi_{\nu\rho} u_\sigma, &&&&l^{\mu}=\epsilon^{\mu\nu\rho\sigma}w_\nu\alpha_\rho u_\sigma,
\end{align}
and to express the variables $T_0$, $\omega$ and $R^2$ in terms of the Lorentz invariants:
\begin{align*}
&\beta^2=\frac{1}{\gamma^2T_0^2}, &&\alpha^2=-(\gamma^2-1)\frac{\omega^2}{T_0^2} 
&&&w^2=-\gamma^2\frac{\omega^2}{T_0^2}, &&&&l^2=-\alpha^2w^2.
\end{align*}
Also, we can decompose the Belinfante stress-energy tensor along the tetrad \eqref{tetrad}, 
with eleven scalar thermodynamic coefficients \footnote{There is a redundancy due to the fact that on the 
rotation axis the defined tetrad is no longer a basis, see the discussion in \cite{ScalarField}.} 
as follows:
\begin{equation}\label{general stress energy tesor}
\begin{split}
T^{\mu\nu}_B(x)=&\rho\, u^\mu u^\nu\! -p\,\Delta^{\mu\nu}\! +W \, w^\mu w^\nu\! +A\,\alpha^\mu \alpha^\nu
    \!+G^l\, l^\mu l^\nu +G\!\left(l^\mu u^\nu\!+l^\nu u^\mu\right)
    \!+\mathbb{A}\!\left(\alpha^\mu u^\nu\!+\alpha^\nu u^\mu\right)\\
&+G^\alpha\!\left(l^\mu \alpha^\nu\!+l^\nu \alpha^\mu\right)
\!+\mathbb{W}\!\left(w^\mu u^\nu\!+w^\nu u^\mu\right)
    \!+A^w\! \left(\alpha^\mu w^\nu\!+\alpha^\nu w^\mu\right)
    \!+G^w\!\left(l^\mu w^\nu\!+l^\nu w^\mu\right).
\end{split}
\end{equation}
To unambiguously identify all terms, we first calculate:
\begin{equation*}
    T^{\mu\nu}_B \frac{l_\nu}{l^2}\frac{l_\mu}{l^2}=G^l l^2-p=\frac{17 \alpha ^4}{2880 \pi ^2 \beta ^4}+\frac{\alpha^2}{72 \beta ^4}-\frac{7 \pi ^2}{180 \beta^4}
   -\frac{w^4}{192 \pi ^2 \beta ^4}+\frac{11 w^2\alpha^2}{160 \pi ^2 \beta ^4}
   +\frac{w^2}{24 \beta^4},
\end{equation*}
and, taking into account that of $l^2=-\alpha^2 w^2$, we can single out $G^l$ and $p$. The remaining 
terms are easily identified:
\begin{equation}\label{coeffrot}
\begin{split}
\rho=& \frac{7\pi^2}{60\beta^4} - \frac{\alpha^2}{24\beta^4} - \frac{w^2}{8\beta^4}
    -\frac{17\alpha^4}{960\pi^2\beta^4} + \frac{w^4}{64\pi^2\beta^4}
   +\frac{23 \alpha^2w^2}{1440\pi^2\beta^4},\\
p=& \frac{7 \pi ^2}{180 \beta ^4}-\frac{\alpha^2}{72 \beta ^4}-\frac{w^2}{24 \beta ^4}-\frac{17 \alpha^4}{2880 \pi ^2 \beta ^4}+\frac{w^4}{192
   \pi ^2 \beta ^4},\\
G^l=& -\frac{11}{160\pi^2\beta^4},\\
G=& \frac{1}{18 \beta ^4}-\frac{31 \alpha^2}{360 \pi ^2 \beta ^4}-\frac{13 w^2}{120 \pi ^2 \beta ^4},\\
W=& -\frac{61\alpha^2}{1440\pi^2\beta^4},\\
A=& -\frac{61 w^2}{1440\pi^2\beta^4},\\
\mathbb{A}=&\mathbb{W}=G^\alpha=G^w=A^w=0.
\end{split}
\end{equation}
The above coefficients coincide with those calculated
in~\cite{Ambrus:2014uqa,ambrus2019exact,tesidiAmbrus} by solving the Dirac equation in cylindrical
coordinates.
In ref.~\cite{tesidiAmbrus}, series similar to those in eq.~\eqref{belinfante 00 example}
and in the appendix \ref{series rotation} have been studied, but a different regularization scheme
has been used.
We stress that, like for the scalar field~\cite{ScalarField}, after setting $w^\mu=0$, the results~\eqref{coeffrot} reduce to the~\eqref{eq:SETCoeffAcc} of equilibrium with pure acceleration. 

We conclude this section by reporting the exact mean value of the axial current. For a massless 
field the axial current is given by the imaginary thermal vorticity series~\eqref{eq:AxialMassless}. 
The analytic distillation and the subsequent analytic continuation can be carried out the same
way as for the series~\eqref{belinfante 00 example} (see appendix \ref{series rotation} for details),
and one obtains:
\begin{equation} \label{eq:JARot}
j\subs{A}^\mu=\frac{1}{\beta^2}\left(\frac{1}{6}
    -\frac{w^2}{24\pi^2}-\frac{\alpha^2}{8\pi^2}\right)\frac{w^\mu}{\sqrt{\beta^2}}.
\end{equation}
Again, this result coincides with previous results evaluated with both 
perturbative~\cite{Buzzegoli:2017cqy,Buzzegoli:2018wpy,Prokhorov:2018bql} and 
exact~\cite{Ambrus:2014uqa,ambrus2019exact,Vilenkin:1980zv} methods. Note that the time-reversal 
symmetry prevents the axial current to be directed along the acceleration vector. Instead, the 
induced axial current along the vorticity of the system is an effect which became known as Axial 
Vortical Effect~\cite{Kharzeev:2015znc}. This effect is allowed by the symmetry of the density
operator and is expected for both massive~\cite{Buzzegoli:2017cqy} and massless Dirac fields at 
equilibrium. Unlike the electric current induced by the rotation, the axial vortical effect can 
contain terms (see eq.~\eqref{eq:JARot}) independent of the axial and chemical potentials 
and that are not constrained by the chiral anomaly~\cite{Buzzegoli:2020ycf}.

\subsection{Thermodynamic equilibrium with rotation and acceleration}

Having successfully tested the fermionic analytic distillation method by comparing its output 
with the results known in literature and otherwise determined, we can apply it to calculate yet 
unknown expressions. Particularly, the global thermodynamic equilibrium with both rotation 
and acceleration along the $z$-axis has never been explored, as it is difficult to define the 
appropriate curvilinear coordinates and solve the relevant field equations. In fact, the analytic 
distillation method, as we are going to show, is much more convenient. 

We thus proceed to study the thermodynamic equilibrium with both rotation and acceleration 
along $z$-axis for the free Dirac field, extending the results obtained for the scalar field in
ref.~\cite{ScalarField}. This is described by the density operator~\eqref{general} with
a thermal vorticity and temperature given by:
\begin{align*}
\varpi_{\mu\nu}=\frac{\omega}{T_0}(g_{\mu 1}g_{\nu 2}-g_{\nu 1}g_{\mu 2})+
\frac{a}{T_0}(g_{\mu 3}g_{\nu 0}-g_{\nu 3}g_{\mu 0}) &&\beta^{\mu}=\frac{1}{T_0}(1+az,-\omega y,\omega x, at).
\end{align*}
Defining once again the tetrad $\{u^\mu,\alpha^\mu,w^\mu,l^\mu\}$ as in eq.~\eqref{tetrad}, 
we first observe that, unlike the previous cases, the scalar $\alpha\cdot w=- a\omega/T_0^2$ is 
non-vanishing and can thus appear in the mean values.

As a case study, we determine the full Belinfante stress-energy tensor for massless fermions.
The tilde-transformed four-temperature vector reads:
\begin{equation*}
    \begin{split}
       n\widetilde{\beta}_n= &\left(i t (\cos (\tfrac{a n}{T_0})-1)+\frac{(1+az)\sin \left(\tfrac{a n}{T_0}\right)}{a} ,-y \sinh
   \left(\tfrac{n \omega }{T_0}\right)+i x \left(\cosh \left(\tfrac{n \omega }{T_0}\right)-1\right),\right.\\
   &\left.x \sinh \left(\frac{n \omega
   }{T_0}\right)+i y \left(\cosh \left(\tfrac{n \omega }{T_0}\right)-1\right),t \sin \left(\tfrac{a n}{T_0}\right)+\frac{i (1+a
   z) \left(\cos \left(\tfrac{a n}{T_0}\right)-1\right)}{a}\right),
    \end{split}
\end{equation*}
and, after replacing $a/T_0\rightarrow -i\Phi, \omega/T_0\rightarrow -i\phi$ the additional
matrices appearing in the series~\eqref{stress energy tensor formula with betatilde} are:
\begin{equation}
\begin{split}
    \label{eq:matricesRotAcc}
    &\Lambda^{\mu}_{\ \nu}\!=\!\!\left(\begin{matrix}
        \cosh n\Phi & 0 & 0 & \sinh n\Phi\\
        0 &\cos n\phi &-\sin n\phi &0\\
        0 &\sin n\phi &\cos n\phi & 0\\
        \sinh n\Phi &0 &0 &\cosh n\Phi
        \end{matrix}\right), \\
    &S(\Lambda)^n=\left(\begin{matrix}
        &\e^{n(\Phi-i\phi)/2} &0 &0 &0\\
        &0 &\e^{-n(\Phi-i\phi)/2} &0 &0\\
        &0 &0 &\e^{-n(\Phi+i\phi)/2} &0\\
        &0 &0 &0 &\e^{n(\Phi+i\phi)/2}
    \end{matrix}\right),\\
    &A^{\mu\nu}\!=\!\!\left(\begin{matrix}
        4\cos (\tfrac{n\phi}{2})\!\cosh (\tfrac{n\Phi}{2}) & 0 & 0 & -4\cos (\tfrac{n\phi}{2})\! \sinh (\tfrac{n\Phi}{2})\\
        0 &-4\cos (\tfrac{n\phi}{2})\!\cosh (\tfrac{n\Phi}{2}) &4\sin (\tfrac{n\phi}{2})\!\cosh (\tfrac{n\Phi}{2}) &0\\
        0 &-4\sin (\tfrac{n\phi}{2})\!\cosh (\tfrac{n\Phi}{2}) &-4\cos (\tfrac{n\phi}{2})\!\cosh (\tfrac{n\Phi}{2}) & 0\\
        4\cos (\tfrac{n\phi}{2})\! \sinh (\tfrac{n\Phi}{2}) &0 &0 &-4\cos (\tfrac{n\phi}{2})\!\cosh (\tfrac{n\Phi}{2})
        \end{matrix}\right),\\
    &A^{\mu\nu}_5 = \left(\begin{matrix}
        &-4\ii\sin\frac{n\phi}{2}\sinh\frac{n\Phi}{2} &0 &0 &4\ii \sin\frac{n\phi}{2} \cosh\frac{n\Phi}{2}\\
        &0 &4\ii\sin\frac{n\phi}{2}\sinh\frac{n\Phi}{2} &4\ii\cos\frac{n\phi}{2}\sinh\frac{n\Phi}{2} &0\\
        &0 &-4\ii\cos\frac{n\phi}{2}\sinh\frac{n\Phi}{2} &4\ii\sin\frac{n\phi}{2}\sinh\frac{n\Phi}{2} &0\\
        &4\ii \sin\frac{n\phi}{2} \cosh\frac{n\Phi}{2} &0 &0 &4\ii\sin\frac{n\phi}{2}\sinh\frac{n\Phi}{2}
    \end{matrix}\right).
\end{split}
\end{equation}
The analytic distillation of the series is described in appendix~\ref{series acc-rot}. 
The coefficients of the general decomposition~\eqref{general stress energy tesor} turn out to
be:
\begin{equation}
\label{coeffaccrot}
 \begin{split}
&\rho= \frac{7\pi^2}{60\beta^4} - \frac{\alpha^2}{24\beta^4} - \frac{w^2}{8\beta^4}
    -\frac{17\alpha^4}{960\pi^2\beta^4} + \frac{w^4}{64\pi^2\beta^4}
    +\frac{23 \alpha^2w^2}{1440\pi^2\beta^4}+\frac{11(\alpha\cdot w)^2}{720\pi^2\beta^4},\\
&p= \frac{7 \pi ^2}{180 \beta ^4}-\frac{\alpha^2}{72 \beta ^4}
    -\frac{w^2}{24 \beta ^4}-\frac{17 \alpha^4}{2880 \pi ^2 \beta ^4}
    +\frac{w^4}{192 \pi ^2 \beta ^4}+\frac{(\alpha\cdot w)^2}{96\pi^2\beta^4},\\
&G^l= -\frac{11}{160\pi^2\beta^4},\\
&G= \frac{1}{18 \beta ^4}-\frac{31 \alpha^2}{360 \pi ^2 \beta ^4}-\frac{13 w^2}{120 \pi ^2 \beta ^4},\\
&W= -\frac{61\alpha^2}{1440\pi^2\beta^4},\\
&A= -\frac{61 w^2}{1440\pi^2\beta^4},\\
&A^w=\frac{61 \alpha\cdot w}{1440\pi^2\beta^4},\\
&\mathbb{A}=\mathbb{W}=G^\alpha=G^w=0.
\end{split}
\end{equation}
These coefficients are consistent with those in eq.~\eqref{coeffrot}, with regard to their
dependence on the scalars $\alpha^2$ and $w^2$. The actual payoff of this case is to have
found out the dependence of the coefficients on the scalar $\alpha\cdot w$, which was vanishing 
in both the pure rotation and the pure acceleration cases. 
We have also calculated the mean axial current, whose expression coincides with \eqref{eq:JARot} 
with no extra dependence on $\alpha\cdot w$ (see appendix~\ref{series acc-rot}).

\section{Massless particles and the chiral kinetic theory}
\label{sec:CKT}

Over the last decade, there has been considerable interest in the relativistic kinetic theory of massless 
fermions, known as chiral kinetic theory~\cite{Son:2012zy,Stephanov:2012ki,Chen:2014cla}. The common approach to this
problem is based on a semi-classical expansion in $\hbar$ of the Wigner equation and an educated {\em ansatz}
of the equilibrium distribution function. In this section, after reviewing the definition of the distribution
function, we obtain exact expressions at global thermodynamic equilibrium.

Because of the decoupling between left and right currents, we can rewrite the particle term of the mean 
current \eqref{jpart} for massless particles as:
\be\label{jpartmassless}
 j^\mu_+(x) = \frac{1}{(2\pi)^3}\sum_{\lambda}\int  \frac{\di^3\p}{2\varepsilon}
 \frac{\di^3\p'}{2\varepsilon'} \, \e^{\ii (p'-p)\cdot x}\langle \wad{p'}{\lambda} \wa{p}{\lambda}\rangle \, 
  \bar{u}_\lambda(p')\gamma^\mu u_\lambda(p)
\ee
where $\lambda$ is the helicity, taking values $-1/2$ and $1/2$. The spinorial product can be calculated 
(see appendix \ref{Spinorial representation}), yielding:
\be\label{uproducts}
 \bar{u}_\lambda(p')\gamma^\mu u_\lambda(p) = \frac{1}{\sqrt{p \cdot \mathfrak{\bar p}\, p^\prime \cdot \mathfrak{\bar p}}}
 \left( p^\mu p^\prime \cdot \mathfrak{\bar p} + p^{\prime\mu} p \cdot \mathfrak{\bar p} 
 - p \cdot p^\prime \mathfrak{\bar p}^\mu + 2 \ii \lambda \epsilon^{\mu\rho\sigma\tau} p^\prime_\rho
  p_\sigma \mathfrak{\bar p}_\tau \right).
\ee  
Instead of replacing the above expression in the \eqref{jpartmassless}, we try to identify a projection
of the right hand side of the equation \eqref{uproducts} onto the four-vector $p$, in accordance to the
methods of chiral kinetic theory. We first observe that $L^\mu = \bar{u}_\lambda(p')\gamma^\mu u_\lambda(p)$
is a complex light-like vector, just by squaring the right hand side of \eqref{uproducts}. Therefore, it can
be decomposed onto two light-like vectors $p$ and $q$ such that $p \cdot q \ne 0$, and a third vector
$N$ which is orthogonal to both $p$ and $q$:
$$
   L^\mu = \frac{L \cdot q}{q \cdot p} p^\mu + \frac{L \cdot p}{q \cdot p} q^\mu + N^\mu(p,q,L)
$$
(this can be taken as a definition of $N$). Now, because of the Dirac equation, $L \cdot p = 0$, and
we are left with:
$$
 \bar{u}_\lambda(p')\gamma^\mu u_\lambda(p) = \frac{\bar{u}_\lambda(p') \slashed q u_\lambda(p)}{q \cdot p} 
 p^\mu + N^\mu .
$$
Using the above decomposition into the current \eqref{jpartmassless} we obtain:
\begin{align*}
 j^\mu_+(x) =& \frac{1}{(2\pi)^3}\sum_{\lambda} \int \frac{\di^3\p}{\varepsilon} p^\mu \int
  \frac{\di^3\p'}{2\varepsilon'} \frac{\bar{u}_\lambda(p') \slashed q u_\lambda(p)}{2 q \cdot p} 
  \, \e^{\ii (p'-p)\cdot x} \langle \wad{p'}{\lambda} \wa{p}{\lambda}\rangle \\ \nonumber
 & + \frac{1}{(2\pi)^3}\sum_{\lambda} \int \frac{\di^3\p}{\varepsilon} N^\mu \int
  \frac{\di^3\p'}{4\varepsilon'} \, \e^{\ii (p'-p)\cdot x} \langle \wad{p'}{\lambda} 
  \wa{p}{\lambda}\rangle.
\end{align*}
The above equation defines a decomposition of the current in phase space ${\cal J}^\mu(p,x)$ (see equation
\eqref{phspacecurrent}) onto the on-shell light-like vector $p$ and a vector $N$ which is perpendicular 
to it. To achieve it, we had to introduce an arbitrary light-like vector $q$; while the decomposition is
inevitably dependent on it, the current is altogether independent thereof once all terms are included.
Following chiral kinetic theory, the decomposition \eqref{jpartmassless} allows to define an on-shell 
$q$-dependent distribution function $f_\lambda(x,p)_{(q)}$ as the coefficient of $p^\mu$ in the decomposition of 
${\cal J}^\mu(p,x)$:
\be\label{chiraldistf}
 f_\lambda(x,p)_{(q)} \equiv \frac{1}{(2\pi)^3} \int
  \frac{\di^3\p'}{2\varepsilon'} \, \e^{\ii (p'-p)\cdot x} \langle \wad{p'}{\lambda} \wa{p}{\lambda}\rangle
  \frac{\bar{u}_\lambda(p') \slashed q u_\lambda(p)}{2 q \cdot p} .
\ee
A suitable choice of $q$ is just the standard vector $\mathfrak{p}$; in this case, the \eqref{chiraldistf}
becomes, by using the \eqref{uproducts}:
$$
 f_\lambda(x,p)_{(\mathfrak{p})} = \frac{1}{(2\pi)^3} \int
  \frac{\di^3\p'}{2\varepsilon'} \sqrt{\frac{p^\prime \cdot \mathfrak{\bar p}}{p \cdot \mathfrak{\bar p}}}
  \, \e^{\ii (p'-p)\cdot x} \langle \wad{p'}{\lambda} \wa{p}{\lambda}\rangle .
$$

The form \eqref{chiraldistf} is especially suitable to obtain an exact expression. Using trace cyclicity
we can write $\bar{u}_\lambda(p') \slashed q u_\lambda(p)=\tr(\slashed{q}u_\lambda(p)\bar{u}_\lambda(p'))$. 
Dealing with massless spinors, we can isolate the helicity lambda using the chiral projectors:
\begin{equation*}
\begin{split}
f_\lambda(x,p)_{(q)} =& \frac{1}{(2\pi)^3} \int
  \frac{\di^3\p'}{2\varepsilon'} \, \e^{\ii (p'-p)\cdot x} \langle \wad{p'}{\lambda} \wa{p}{\lambda}\rangle
  \frac{\tr( \slashed q u_\lambda(p)\bar{u}_\lambda(p'))}{2 q \cdot p} \\
  =&\frac{1}{(2\pi)^3} \int
  \frac{\di^3\p'}{2\varepsilon'} \frac{\e^{\ii (p'-p)\cdot x}}{2 q \cdot p}\sum_{s,t}
  \tr\left( \slashed q \frac{I+2\lambda\gamma_5}{2}\langle \wad{p'}{t} \wa{p}{s}\rangle 
  u_s(p)\bar{u}_t(p')\frac{I-2\lambda\gamma_5}{2}\right)\\
  =&\frac{1}{(2\pi)^3}\frac{1}{2 q \cdot p}\tr\left( \slashed q \frac{I+2\lambda\gamma_5}{2} \int
  \frac{\di^3\p'}{2\varepsilon'} \, \e^{\ii (p'-p)\cdot x} \sum_{s,t}
  \langle \wad{p'}{t} \wa{p}{s}\rangle u_s(p)\bar{u}_t(p')\right),
 \end{split}
\end{equation*}
where we used the anticommutation rules of $\gamma_5$ and  $\gamma_5 u_\lambda(p)=2\lambda u_\lambda(p)$, 
where $\lambda$ is the helicity and $2\lambda$ the chirality. Also note that $(2\lambda)^2=1$.

Except for $\slashed{q}$ and the chiral projector, the expression in the trace, and in particular the 
product $\langle \wad{p'}{t} \wa{p}{s}\rangle u_s(p)\bar{u}_t(p')$, is very similar to the particle 
part of the Wigner function \eqref{Wigner function expanded}. The substitution of the exact result 
\eqref{number operator iterative result} and the subsequent calculations can be done as in section 
\ref{subsecc: calculations exact Wigner}, and one gets: 
$$\int
  \frac{\di^3\p'}{2\varepsilon'} \, \e^{\ii (p'-p)\cdot x} \sum_{s,t}
  \langle \wad{p'}{t} \wa{p}{s}\rangle u_s(p)\bar{u}_t(p')=\sum_{n=1}^{\infty}(-1)^{n+1}
  \e^{-n\widetilde{\beta}_n\cdot p}S(\Lambda)^n(\slashed{p}+m).
$$
Plugging this result in the trace above, we have:
$$
    f_\lambda(x,p)_{(q)} = 
    \frac{1}{(2\pi)^3}\frac{1}{2 p\cdot q}\sum_{n=1}(-1)^{n+1}\e^{-n\widetilde{\beta}_n\cdot p}
    \tr\left(\frac{I+2\lambda\gamma_5}{2} S(\Lambda)^n\slashed{p}\slashed{q}\right).
$$
Expanding the $\slashed{p}\slashed{q}$ product, we obtain the final expression of the exact
equilibrium distribution function:
\begin{align}\label{exactchiralf}
f_\lambda(x,p)_{(q)} &= {\rm dist}_{\varpi=0} 
   \sum_{n=1}^{\infty}\frac{(-1)^{n+1}}{2(2\pi)^3}\e^{- n \widetilde{\beta}(-n\varpi)\cdot p}
  \left\{ \tr\left(\frac{I+2\lambda\gamma_5}{2}\exp\left[ n \frac{\varpi_{\rho\sigma}}{2} \Sigma^{\rho\sigma} \right] 
  \right) \right.  \\ \nonumber
    & \left. + \frac{2\ii q_\mu p_\nu}{q\cdot p} \tr\left(\frac{I+2\lambda\gamma_5}
    {2}\Sigma^{\mu\nu} \exp \left[ n \frac{\varpi_{\rho\sigma}}{2} \Sigma^{\rho\sigma}\right] \right) \right\},
\end{align}
where the real thermal vorticity has been introduced in the series, provided that it is made 
convergent by the analytic distillation in $\varpi=0$. This expression gives rise to a current 
term along $p^\mu$ which is in agreement with the general expression \eqref{equilibrium particle current}.

Applying analytic distillation to the \eqref{exactchiralf} goes beyond the scope of this work.
Indeed, the theorem \ref{th1} cannot be applied straightforwardly as the series is not of the required 
functional form, so a new method to obtain the full asymptotic power series in $\varpi$ is needed. 
The \eqref{exactchiralf} differs from the usual {\em ansatz} of the equilibrium distribution 
\cite{Chen:2014cla,Liu:2018xip,Liu:2020flb,Shi:2020htn}
mostly for the exponential factor and for the absence of the frame vector.

\section{Spin density matrix and spin polarization vector}
\label{sec:spindensity}

As has been mentioned in the introduction, there is a phenomenological considerable interest on spin 
and polarization in relativistic fluids ~\cite{Becattini:2020ngo}. In the context of heavy ion collisions, the vorticity of the quark gluon plasma is transferred to the quasi free fermions (e.g. the $\Lambda$ particles) during hadronization, resulting in a non-vanishing polarization of hadrons. Our method provides the opportunity 
to calculate an exact expression of the spin polarization vector and the spin density matrix at global 
equilibrium with acceleration and rotation. Such expressions are still unknown and the formulae used in 
literature are just the leading order term in the thermal vorticity.  Even though thermal vorticity is 
a small quantity in all phenomenological applications, it is important to know the exact expression
to estimate the quantitative impact of higher order terms.

The spin density matrix - for a free field - is defined in a quantum field theoretical framework as:
\begin{equation*}
 \Theta_{rs}(p)=\frac{\Tr\left(\wrho\, \wad{p}{s} \, \wa{p}{r}\right)}
 {\sum_t\Tr\left(\wrho\, \wad{p}{t} \, \wa{p}{t}\right)},
\end{equation*} 
where $r,s,t$ are the spin states labels; this matrix is the full description of the spin state 
of a particle. For massive particles, the spin density matrix and the spin polarization vector can
be expressed by means of the Wigner function as \cite{becattini2020polarization}:
\begin{align}
\Theta(p)=&\frac{\int \di\Sigma\cdot p \; \bar{U}(p) W_+(x,p) U(p)}
    {\tr_2 \left(\int \di\Sigma\cdot p \; \bar{U}(p)W_+(x,p) U (p)\right)},\label{spindensitymatrix}  \\ 
S^{\mu}(p) =&\frac{1}{2}\frac{\int \di\Sigma\cdot p \; \tr(\gamma^\mu \gamma_5 W_+(x,p))}
    {\int \di\Sigma\cdot p \; \tr(W_+(x,p))}\label{spinpolvect},
\end{align}
where the integration over the (arbitrary) space-like hypersurface $\Sigma$ with the measure 
$\di \Sigma \cdot p$ makes the momentum $p$ on-shell. This is a general feature for free fields 
\cite{becattini2020polarization} and we are going to explicitly prove it for the Wigner function 
$W(x,k)$ found in \eqref{exact Wigner function}. First of all, for any space-like hypersurface $\Sigma$, 
one has:
\be\label{vanishing}
 \int_\Sigma \di \Sigma \cdot k \partial_\mu W_\pm (x,k) = 0 .
\ee
To show it, one should keep in mind that from \eqref{tildebetax} one has
$$
 -n\widetilde{\beta}(-n\varpi)\cdot p = -n\widetilde{b}(-n\varpi)\cdot p -\ii x\cdot(\Lambda^n p-p),
$$
so that, taking the derivatives of the exponent in \eqref{exact Wigner function} the
following factors are obtained:
$$
 k \cdot ((\Lambda^n p-p)) \delta^4\left( k-\frac{\Lambda^n p+p}{2} \right), \qquad \qquad
 k \cdot ((\Lambda^n p-p)) \delta^4\left( k+\frac{\Lambda^n p+p}{2} \right),
$$
which both vanish, proving the \eqref{vanishing}. Now, provided that suitable boundary conditions 
apply, we can then calculate the integral:
$$
  \int_\Sigma \di \Sigma_\mu k^\mu W_+(x,k)
$$  
over any hypersurface, and particularly the hypersurface $t=0$. We thus obtain, by integrating the
\eqref{exact Wigner function} in $\di^3 x$, a factor $\delta^3({\bf k} - {\bf p})$ in the particle 
term of the Wigner function, hence:
\be\label{Wignerint}
  \int_\Sigma \di \Sigma_\mu k^\mu W_+ (x,k) = \delta(k^0 - \varepsilon_{k}) \frac{1}{2} 
  \sum_{n=1}^{\infty}(-1)^{n+1} \e^{-n\widetilde{b}(-n\varpi) \cdot k} S(\Lambda)^n (m+\slashed{k}) 
\ee  
which makes $k$ manifestly on shell. 

Therefore, by using the \eqref{Wignerint} with $k=p$, and continuing the Wigner function to real
thermal vorticity (with associated distillation) the \eqref{spindensitymatrix} gives rise to:
\be\label{spindensityfinal}
 \Theta(p)={\rm dist}_{\varpi=0}\frac{\sum_{n=1}^{\infty}(-1)^{n+1} \e^{-n\widetilde{b}(-n\varpi) \cdot p} 
\bar U(p) \exp [n\varpi:\Sigma/2] U(p)}{\tr_2 \sum_{n=1}^{\infty}(-1)^{n+1} \e^{-n\widetilde{b}(-n\varpi) \cdot p} 
\bar U(p) \exp [n\varpi:\Sigma/2] U(p)},
\ee
where the Dirac equation for the spinors $U$ has been used. Similarly, the spin polarization 
vector \eqref{spinpolvect} at global thermodynamic equilibrium becomes:
\be\label{spinvectorfinal}
\begin{split}
    S^\mu(p)={\rm dist}_{\varpi=0}\frac{1}{2m}\frac{\sum_{n=1}^{\infty}(-1)^{n+1} \e^{-n\widetilde{b}(-n\varpi) \cdot p} 
    \tr\left( \gamma^\mu\gamma_5 \exp [n\varpi:\Sigma/2] \slashed{p}\right)}
    { \sum_{n=1}^{\infty}(-1)^{n+1} \e^{-n\widetilde{b}(-n\varpi) \cdot p} \tr\left(\exp [n\varpi:\Sigma/2]\right)}\\
    ={\rm dist}_{\varpi=0}\frac{1}{2m}\frac{\sum_{n=1}^{\infty}(-1)^{n+1} \e^{-n\widetilde{b}(-n\varpi) \cdot p} 
    p_\nu A^{\mu\nu}_5(n,\varpi)}
    { \sum_{n=1}^{\infty}(-1)^{n+1} \e^{-n\widetilde{b}(-n\varpi) \cdot p} \tr\left(\exp [n\varpi:\Sigma/2]\right)},
    \end{split}
\ee
where we used \eqref{definition A5} for real thermal vorticity. For the global equilibrium with pure 
acceleration, described in the subsection \ref{subsec:acc}, we have:
$$
\e^{n\varpi:\Sigma/2} = \left(\begin{matrix}
    &\e^{\ii n a/2 T_0} &0 & 0 & 0\\
    &0 &\e^{-\ii n a/2 T_0} &0 & 0\\
    &0 &0 &\e^{-\ii n a/2 T_0} &0\\
    &0 &0 &0 &\e^{\ii n a /2 T_0}\\
    \end{matrix}\right) \quad  
    A^{\mu\nu}_5=\left(\begin{matrix}
    &0 &0 &0 &0\\
    &0 &0 &- 4 \sin \frac{n a}{2 T_0} &0\\
    &0 & 4 \sin \frac{n a}{2 T_0} &0 &0\\
    &0 &0 &0 &0
    \end{matrix}\right).
$$
Hence, at the Unruh temperature $T_0 = a/2\pi$, the first matrix becomes the identity and the tensor 
$A_5$ vanishes, making the spin polarization vector in eq.~\eqref{spinvectorfinal} zero. This is in 
excellent agreement with the expectations from the physics of the Unruh effect, which implies the 
equivalence of the Minkowski vacuum (with, of course, no polarization) at the finite temperature 
$a/2\pi$ for an accelerated observer. 

As discussed at the end of section~\ref{sec:currents}, it is possible to obtain a finite full 
expression at linear order in real thermal vorticity by expanding in $\varpi$:
$$
\tr\left(\gamma^\mu \gamma_5 S(\Lambda)^n\slashed{p}\right)\sim-n\varpi_{\alpha\beta}p_\nu
\epsilon^{\mu\alpha\beta\nu}+{\cal O}(\varpi^2), \qquad \qquad 
\e^{-n\widetilde{b}(-n\varpi)\cdot p}\sim e^{-b\cdot p}+{\cal O}(\varpi),
$$
and summing the series in $n$. The result is:
\begin{equation}\label{linearappr}
S^{\mu}(p)\sim -\frac{1}{8m} \epsilon^{\mu\alpha\beta\nu}\varpi_{\alpha\beta}p_\nu 
 \frac{n_F(b\cdot p)(1-n_F(b\cdot p))}{n_F(b\cdot p)},
\end{equation}
where $n_F$ is the Fermi-Dirac distribution function. This formula is in full agreement with the local 
equilibrium expression found in ref.~\cite{Becattini:2013fla}. We leave the study of the full resummation 
of the series \eqref{spinvectorfinal} to future work.

\section{Summary}
\label{summary}

In summary, we have derived a general exact form of the Wigner function and the thermal expectation
values of local operators of the free Dirac field in the most general case of global thermodynamic 
equilibrium in Minkowski space-time, that is with a Killing four-temperature vector including rotation 
and acceleration. Our method is an extension of that used for the scalar field in ref.~\cite{ScalarField}:
a factorization of the density operator \eqref{general} and the derivation of the thermal expectation 
values of quadratic combinations of creation and annihilation operators by iteration.
For the spin 1/2 particles, we have obtained the general form of the Wigner function of a free Dirac
field as a formal series for imaginary thermal vorticity including all quantum corrections to the 
classical term. The analytic continuation to real thermal vorticity and the extraction of finite results, 
demands the application of the analytic distillation, an operation on complex functions introduced in 
ref.~\cite{ScalarField} and extended here to the alternate fermionic series. We find that, in the pure 
acceleration massless case, the method of analytic distillation leads to expressions which all vanish 
at the Unruh temperature, in agreement with previous findings. Similarly to the scalar field case, 
analytic distillation defines a new class of complex polynomials which all vanish at $z=2\pi \ii$.

We have studied the series in two major cases of non-trivial equilibrium, the pure acceleration and the
pure rotation and compared with known perturbative and exact results obtained solving Dirac equation in
rotating coordinates. A new exact result has been obtained, namely the thermal expectation
value of the axial current and stress-energy tensor in the case of global equilibrium with both
acceleration and rotation for the massless case. We have derived the exact expression of the distribution 
function for massless fermions at global thermodynamic equilibrium, which is an important result for the 
chiral kinetic theory.

Finally, we derived the exact form of the spin density matrix and the spin polarization vector for
the spin 1/2 particles at global thermodynamic equilibrium as a function of thermal vorticity.
 This latter result is applicable as an improved approximation, with respect to 
\eqref{linearappr} to situations where the fields are quasi-free, for instance the final state
baryons in relativistic heavy ion collisions after hadronization.

\acknowledgments
M.B. is supported by the Florence University fellowship ``Effetti quantistici nei fluidi
relativistici''.


\appendix
\section{Spinors and group theory}
\label{Spinorial representation}

We review the spinor formalism from a group theory viewpoint. Here, we will focus on 
the spin $1/2$ case, but the formalism can be extended to any spin \cite{Weinberg:1964cn,Weinberg:1964ev}. 
As it is well known, the Dirac field describes a particle (as well as an antiparticle) of spin $1/2$, 
and it transforms in the $(0,1/2)\oplus(1/2,0)$ projective representation of the ortochronous Lorentz 
group ${\rm SO}(1,3)^\uparrow$. 

The one-to-one correspondence between four-vectors and $2\times 2$ hermitian matrices is defined
through: 
\begin{equation*}
    \unt{x}=x^\mu\sigma_\mu=x^0\mathbb{I}+x^1\sigma_1+x^2\sigma_2+x^3\sigma_3=\left(\begin{matrix}
        x^0+x^3 &x^1-ix^2\\
        x^1+ix^2 &x^0-x^3
    \end{matrix}\right).
\end{equation*}  
where $\sigma_\mu=(\mathbb{I},\sigma_1,\sigma_2,\sigma_3)$ and $\sigma_i$ are the Pauli matrices; 
also the notation $\bar{\sigma}_\mu=(\mathbb{I},-\bm{\sigma})$ will be used. 
The Lorentz transformations $\Lambda$ are represented by a ${\rm SL}(2,\mathbb{C})$ matrix~$D(\Lambda)$:
\begin{equation*}
\unt{\Lambda x}=D(\Lambda) \unt{x} D(\Lambda)^\dagger ,
\end{equation*} 
The complex matrix $D(\Lambda)$ is determined up to a sign, so that the ${\rm SL}(2,\mathbb{C}) 
\leftrightarrow {\rm SO}(1,3)^\uparrow$ is a 2 to 1 correspondence. The above definition identifies 
the $(0,1/2)$ projective representation $D^{(0,1/2)}$ of  ${\rm SO}(1,3)^\uparrow$, where 
the generators of boosts and rotations are respectively $D^{(0,1/2)}(K_i)=i\sigma_i/2$ 
and $D^{(0,1/2)}(J_i)=\sigma_i/2$. Therefore:
$$
 D^{(0,1/2)}(\Lambda) = D(\Lambda)
$$
We also define the following map:
\begin{equation*}
    \upt{x} \equiv x^\mu\bar{\sigma}_\mu=x^0\mathbb{I}-x^1\sigma_1-x^2\sigma_2-x^3\sigma_3=
    \left(\begin{matrix} x^0-x^3 &-x^1+ix^2\\
        -x^1-ix^2 &x^0+x^3
    \end{matrix}\right).
\end{equation*}
It is easy to show that:
\begin{equation}\label{upt unt product}
    \upt{x}\unt{x}=\unt{x}\upt{x}= x \cdot x \mathbb{I} = x^2\mathbb{I},
\end{equation}
Since \eqref{upt unt product} is a Lorentz invariant, we can infer the transformation law of 
$\upt{x}$ from that of $\unt{x}$. For the eq.~\eqref{upt unt product} to be valid in any frame, 
one  needs:
\begin{equation*}
    \upt{\Lambda x}=D(\Lambda)^{\dagger-1} \upt{x} D(\Lambda)^{-1} .
\end{equation*}
This map actually corresponds to the $(1/2,0)$ representation of ${\rm SO}(1,3)^\uparrow$, whose 
generators are $D^{(1/2,0)}(K_i)=-i\sigma_i/2$ and $D^{(1/2,0)}(J_i)=\sigma_i/2$. Hence:
$$
D^{(1/2,0)}(\Lambda)=D^{(0,1/2)}{(\Lambda)^{\dagger}}^{-1} = D(\Lambda)^{\dagger -1}
$$ 

The quantum single particles states of momentum $p$ in the Hilbert space are usually defined starting 
from a standard four-momentum $\mathfrak{p}$, and using a standard Lorentz transformation $[p]$ 
(depending on $p$ and $\mathfrak{p}$) such that $\mathfrak{p}\mapsto p$. In formulae:
\begin{equation}\label{transformations in spiorial rep}
    D([p])\unt{\mathfrak{p}}D([p])^{\dagger}=\unt{p}, \qquad \qquad 
    {D([p])^{\dagger}}^{-1}\upt{\mathfrak{p}}D([p])^{-1}=\upt{p}.
\end{equation}
Also notice, from \eqref{upt unt product}, $\unt{p}\upt{p}=m^2 \mathbb{I}$. The choice of 
$\mathfrak{p}$, which is in principle arbitrary, requires a separate handling of massive and 
massless particles. Indeed, for massive particles $\mathfrak{p}=(m,0)$ is basically the only
option, whereas in the massless case one usually chooses $\mathfrak{p}=(\kappa,0,0,\kappa)$ 
for some positive energy $\kappa$.  
A crucial role in the Lorentz transformation rules of creation and annihilation operators, 
and therefore of the field, is played by the so-called \emph{little group} of $\mathfrak{p}$, that is the 
group of transformations leaving $\mathfrak{p}$ invariant. In the case of massive particles with 
$\mathfrak{p}=(m,0)$, we have $\upt{\mathfrak{p}}=\unt{\mathfrak{p}}=m\mathbb{I}$. Requiring the 
invariance of these matrices in the equations \eqref{transformations in spiorial rep}, it turns out
that the little group is the rotation group SO(3). In the case of massless particles, choosing the 
standard vector as $\mathfrak{p}=(\kappa,0,0,\kappa)$, we have:
$$
\unt{p}=\left(\begin{matrix}&2\kappa &0\\
&0 & 0\end{matrix}\right),\qquad \qquad \unt{p}=\left(\begin{matrix}&0 &0\\
&0 & 2\kappa\end{matrix}\right),
$$
and the transformations of the little group must have the form:
\begin{equation*}
D([\mathfrak{p}])=\left(\begin{matrix}&\e^{-i\phi/2} &Z\e^{-i\phi/2}\\
&0 & \e^{i\phi/2}\end{matrix}\right).
\end{equation*}
The parameter $\phi$ is associated with a rotation around the $z$ axis and $Z$ (a complex number) 
corresponds to a translation in the Euclidean plane \cite{MoussaStora}.

It can be realized that $W(\Lambda,p)=[\Lambda p]^{-1}\Lambda [p]$ belongs to the little group of 
$\mathfrak{p}$ because it maps $\mathfrak{p}$ to $p$, then to $\Lambda p$ and finally back to 
$\mathfrak{p}$. The $W(\Lambda,p)$, usually known as the Wigner rotation, dictates how creation 
and annihilation operators transform under Lorentz transformations represented in the Hilbert space:
\begin{equation}\label{trasf a}
    \widehat\Lambda \wa{p}{r} \widehat\Lambda^{-1} = 
    \sum_s D(W(\Lambda,p))^\dagger_{rs}\wa{\Lambda p}{s}.
\end{equation}
As remarked above, if the particle is massive, $W(\Lambda,p)$ is a rotation, hence  $D(W(\Lambda,p))$
is a unitary matrix. In the massless case, it is generally non-unitary, unless the translation parameter 
$Z$ is set to zero. This is precisely the feature of actual physical representations: known massless 
particles states transform in representations with $Z=0$ \cite{tung1985group}. Therefore, altogether, the 
little group in both massive and massless cases always acts on physical states with unitary transformations, 
i.e. $D(W)^{\dagger}=D(W)^{-1}$ in \eqref{trasf a}.

The field in eq.~\eqref{field expansion compact} with the spinors \eqref{spinors2} transforms as the 
$(0,1/2)\oplus(1/2,0)$ representation of the Lorentz group:
\begin{equation*}
\widehat\Lambda \Psi(x) \widehat\Lambda^{-1} = S(\Lambda)^{-1}\Psi(\Lambda x),   
\end{equation*}
where $S(\Lambda)$ is given in the eq.~\eqref{slambda1}. We show this for the particle term only:
\begin{equation*}
    \Psi_+(x)=\frac{1}{(2\pi)^{\frac{3}{2}}}\int\frac{\di^3\p}{2\varepsilon} \e^{-ip\cdot x}U(p)\wA(p).
\end{equation*}
Making use of the transformation rule \eqref{trasf a}, which in compact notation reads:
\begin{equation*}
    \widehat\Lambda \wA(p) \widehat\Lambda^{-1} = D(W(\Lambda,p))^{\dagger} \wA(\Lambda p),
\end{equation*}
the transformation of the field is:
\begin{equation*}
\widehat\Lambda \Psi_+(x) \widehat\Lambda^{-1}=
\frac{1}{(2\pi)^{\frac{3}{2}}}\int\frac{\di^3\p}{2\varepsilon} \e^{-ip\cdot x}U(p)D(W(\Lambda,p))^{\dagger}
\wA(\Lambda p).
\end{equation*}
Plugging the spinor as in \eqref{spinors2} and making use of the invariance of  $\mathfrak{p}$
under $D(W(\Lambda,p))^{\dagger}=D(W(\Lambda,p))^{-1}=[p]^{-1}\Lambda^{-1}[\Lambda p]$, as well
as the transformation rules \eqref{transformations in spiorial rep}, we have:
\begin{equation*}\begin{split}
    &U(p)D(W(\Lambda, p))^{\dagger}=
    \left(\begin{matrix}
    & D([p]) &0\\
    &0 &{D([p])^{\dagger}}^{-1}
    \end{matrix}\right)
    \left(\begin{matrix}\unt{\mathfrak{p}}D(W(\Lambda,p))^{\dagger}\\
    \upt{\mathfrak{p}}D(W(\Lambda,p))^{-1}\end{matrix}\right)=\\
    =& \left(\begin{matrix}D([p])D([p]^{-1}\Lambda^{-1}[\Lambda p])\unt{\mathfrak{p}}\\
    D({[p]^\dagger}^{-1})D([p]^{\dagger}\Lambda^\dagger{[\Lambda p]^\dagger}^{-1})\upt{\mathfrak{p}}\end{matrix}\right)= \left(\begin{matrix}D(\Lambda^{-1}) &0\\
    0 & {D(\Lambda^{-1})^\dagger}^{-1}\end{matrix}\right)U(\Lambda p)=S(\Lambda)^{-1}U(\Lambda p) .
    \end{split}
\end{equation*}
Finally, after changing the integration variable from $p$ to $\Lambda p$, we obtain:
$$
 \widehat\Lambda \Psi_+(x) \widehat\Lambda^{-1} =S(\Lambda)^{-1}\Psi_+(\Lambda x) \; ;
$$
for the antiparticle term the proof is similar. The generators of $S(\Lambda)$ can be written
in the form: 
$$
\Sigma^{\mu\nu}=\left(\begin{matrix}&D^{(0,1/2)}(J^{\mu\nu}) &0\\ & 0 &D^{(1/2,0)}(J^{\mu\nu})\end{matrix}\right).
$$
By using their form in terms of Pauli matrices, it can be readily checked that they coincide
with the better known form $(\ii/4)[\gamma^\mu,\gamma^\nu]$ where $\gamma$ are in the so-called
Weyl representation \eqref{gammas}.

It is also useful to show that the spinors in eq.~\eqref{spinors2} fulfill the Dirac equation. 
For the standard momentum $\mathfrak{p}$, we have:
\begin{equation*}
    \slashed{\mathfrak{p}}U(\mathfrak{p})=N\left(\begin{matrix}0 & \unt{\mathfrak{p}}\\
     \upt{\mathfrak{p}} &0\end{matrix}\right)\left(\begin{matrix}\unt{\mathfrak{p}}\\
    \upt{\mathfrak{p}}\end{matrix}\right)=N m\left(\begin{matrix}m\mathbb{I}\\
    m\mathbb{I}\end{matrix}\right),
\end{equation*}
with $N$ normalization factor. Since in the massive case $\unt{\mathfrak{p}}=\upt{\mathfrak{p}}=m\mathbb{I}$, 
and in the massless case $\upt{\mathfrak{p}}\unt{\mathfrak{p}}=m^2=0$, it is seen that the Dirac 
equation is satisfied for $p=\mathfrak{p}$. From the transformation rules of the spinors:
$$
  U(p)D(W(\Lambda, p))^{\dagger}  = S(\Lambda)^{-1}U(\Lambda p)
$$
it follows that the equation is fulfilled for any $p$ obtained from $\mathfrak{p}$ by means
of a Lorentz transformation. Similarly, one can show that the $V(p)$ spinors fulfill Dirac equation.

The spinors can be written by choosing a particular standard transformations. For instance, in 
the massive case, the standard transformation can be chosen as {\em the} pure boost transforming 
$\mathfrak{p}$ to $p$. This transformation can be written in the $(0,1/2)$ representation as
\cite{MoussaStora}:
\begin{equation*}
    D([p])=\frac{m+\unt{p}}{\sqrt{2m(\varepsilon+m)}}
\end{equation*}
whence the particle spinor in \eqref{spinors2} follows:
\begin{equation*}
    U(p)=\frac{1}{\sqrt{2m(\varepsilon+m)}}\left(\begin{matrix}&m+\unt{p} &0\\
    &0 &m+\upt{p}\end{matrix}\right)\frac{1}{\sqrt{2m}}\left(\begin{matrix}&m\mathbb{I}\\
    &m\mathbb{I}\end{matrix}\right)=\frac{m+\slashed{p}}{\sqrt{2m(\varepsilon+m)}}U(\mathfrak{p}).
\end{equation*}
For massless particles, the standard transformation can be chosen as the boost along the $z$
axis followed by the rotation of axis ${\bf k} \times {\bf p}$ mapping the $z$ axis to the direction 
of $\bf{p}$. Though more involved, it is then possible to show that \cite{DeGroot:1980dk}:
\begin{equation}\label{spinor from standard spinor massless}
    U(p)=\left(\begin{matrix}&D([p]) &0\\
    &0 &{D([p])^{\dagger}}^{-1}\end{matrix}\right)\frac{1}{\sqrt{2\kappa}}\left(\begin{matrix}
    &\unt{\mathfrak{p}}\\&\upt{\mathfrak{p}}\end{matrix}\right)=\frac{\slashed{p}\gamma^0}
    {\sqrt{2 p\cdot \bar{\mathfrak{p}}}}U(\mathfrak{p}) .
\end{equation}
%
\subsection{Massless spinor product}

Taking advantage of the above formulae, it is possible to calculate a useful expression, the spinor 
product $\bar{u}_\lambda(p') \gamma^{\mu} u_\lambda(p)$ for massless particles. To begin with, spinors 
are written in terms of the standard spinor via the eq.~\eqref{spinor from standard spinor massless}:
\begin{equation*}
    \bar{u}_\lambda(p') \gamma^{\mu} u_\lambda(p)=\frac{1}{2\sqrt{(p\cdot\bar{\mathfrak{p}})
    (p'\cdot\bar{\mathfrak{p}})}}\bar{u}_\lambda(\mathfrak{p})\gamma^0\slashed{p'}\gamma^\mu\slashed{p}
    \gamma^0u_\lambda(\mathfrak{p}).
\end{equation*}
The product of matrices in the middle of the right hand side can be written in two ways:
\begin{equation*}
    \slashed{p'}\gamma^\mu\slashed{p}=\left\{\begin{array}{c}
         2p'^{\mu}\slashed{p}-\gamma^{\mu}\slashed{p'}\slashed{p}  \\
        2p^{\mu}\slashed{p'}-\slashed{p'}\slashed{p}\gamma^{\mu}
    \end{array}\right.=\left\{\begin{array}{c}
         2p'^{\mu}\slashed{p}-\gamma^{\mu}p\cdot p'+2i\gamma^{\mu}p'^\rho p^\sigma\Sigma_{\rho\sigma}  \\
        2p^{\mu}\slashed{p'}-\gamma^{\mu}p\cdot p'+2ip'^\rho p^\sigma\Sigma_{\rho\sigma}\gamma^{\mu}
    \end{array}\right. ,
\end{equation*} 
where we used the anticommutation rules of the $\gamma$ matrices and the known expression for 
the product $\slashed{p}\slashed{p'}$. Taking the half-sum of the two:
\begin{equation}\label{intermediate}
    \bar{u}_\lambda(p') \gamma^{\mu} u_\lambda(p)=\frac{\bar{u}_\lambda(\mathfrak{p})\gamma^0
    \left(2p^{\mu}\slashed{p'}+2p'^{\mu}\slashed{p}-2\gamma^{\mu}p\cdot p'+2ip'^\rho 
    p^\sigma\{\Sigma_{\rho\sigma},\gamma^{\mu}\}\right)\gamma^0
    u_\lambda(\mathfrak{p})}{4\sqrt{(p\cdot\bar{\mathfrak{p}})(p'\cdot\bar{\mathfrak{p}})}}.
\end{equation}
This equation can be simplified by using the known relation:
$$
\{\Sigma^{\rho\sigma},\gamma^\mu\}=-\epsilon^{\mu\rho\sigma\tau}\gamma_\tau\gamma_5
$$
as well as $\gamma_5 u_\lambda(p)=2\lambda u_\lambda(p)$. Thereby, all terms in the equation
\eqref{intermediate} involve the calculation of the product:
$$
\bar{u}_\lambda(\mathfrak{p})\gamma^0\gamma^{\alpha}\gamma^0u_\lambda(\mathfrak{p}) .
$$
With $\gamma^0\gamma^{0}\gamma^0=\gamma^0$ and $\gamma^0\gamma^{i}\gamma^0=-\gamma^i$, as well 
as $\bar{u}_\lambda(p)\gamma^{\alpha}u_\lambda(p)=2p^\mu$, this yields:
\begin{equation*}
    \bar{u}_\lambda(\mathfrak{p})\gamma^0\gamma^{\alpha}\gamma^0u_\lambda(\mathfrak{p})=2\bar{\mathfrak{p}}^\alpha,
\end{equation*}
where the bar over a vector implies space reflection, i.e. $\bar{p}=\left(p^0,-\bm{p}\right)$. 
By using the above expression into the \eqref{intermediate}, we finally obtain:
\begin{equation*}
    \bar{u}_\lambda(p') \gamma^{\mu} u_\lambda(p)=\frac{1}{\sqrt{(p\cdot\bar{\mathfrak{p}})
    (p'\cdot\bar{\mathfrak{p}})}}\left(p^{\mu}\bar{\mathfrak{p}}\cdot p'+p'^{\mu}p\cdot \bar{\mathfrak{p}}-
    \bar{\mathfrak{p}}^{\mu}p\cdot p'+2i\lambda p'_\rho p_\sigma \bar{\mathfrak{p}}_\tau\epsilon^{\mu\rho\sigma\tau}\right).
\end{equation*}
%

\section{Massless fermions and Unruh effect}
\label{sec:AccSeries}

In this section we show that the TEVs at global equilibrium with pure acceleration (see subsection 
\ref{subsec:acc}) of a large class of local operators quadratic in the field, vanish at the Unruh 
temperature $T_0= a/2\pi$, just like for the scalar field \cite{ScalarField}. We show it specifically 
for the massless Dirac field.
The TEVs of local operators quadratic in the field are given by an integral of the Wigner function, 
just like the operators in eqs.~\eqref{mean current}, \eqref{mean axial current} and \eqref{meanset}. 
Hence, by using the exact form of the Wigner function~\eqref{exact Wigner function} they are given 
by momentum integrals of this sort:
\begin{equation}\label{eq:momint}
\int\frac{\di^3\p}{\varepsilon}\,p^{\mu_1}\cdots p^{\mu_N}\, \partial^{\nu_1}\cdots \partial^{\nu_M}\,
    \e^{- n \tilde \beta_n \cdot p} = \partial^{\nu_1}\cdots \partial^{\nu_M}\,
    \frac{\partial}{\partial \tilde\beta_n^{\mu_1}}
    \ldots \frac{\partial}{\partial \tilde\beta_n^{\mu_N}}
    \frac{(-1)^N}{n^N} \int\frac{\di^3\p}{\varepsilon} \; \e^{- n \tilde \beta_n \cdot p}.
\end{equation}
These integrals ought to be evaluated for imaginary acceleraration~\eqref{eq:btildeAcc}, 
and then, after analytic distillation, they can be analytically continued.

As dicussed in subsection~\ref{subsec:acc}, it is indeed sufficient to calculate the 
integrals~\eqref{eq:momint} in $x=0$. By using the eq.~\eqref{eq:MasslessInt}
and the four-temperature~\eqref{eq:btildeAcc}, it can be seen that integrals such as 
\eqref{eq:momint}, evaluated in $x=0$, result in linear combinations of the following series:
\begin{equation*}
S_{F,2m+2}(\phi)=\sum_{n=1}^\infty (-1)^{n+1}\frac{\phi^{2m+2}\sinh(n\phi)}{\sinh^{2m+3}(n\phi/2)}.
\end{equation*}
Since $S_{F,2m+2}$ is a uniformly convergent series of analytic functions for ${\rm Re}\,\phi \neq 0$, 
it defines an analytic function therein. However, the series is badly divergent for purely
imaginary $\phi$ and we need to carry out its analytic distillation at $\phi=0$ in order to obtain 
a finite physical thermal expectation value. For this purpose we write the series introducing a 
function $F$ defined by:
\begin{equation*}
S_{F,2m+2}(\phi) = \phi^{2m+2} \sum_{n=1}^\infty (-1)^{n+1} F(n\phi).
\end{equation*}
In this form we can apply the theorem \ref{th1} to obtain an asymptotic series for $S_{F,2m+2}$
once we know an asymptotic power expansion about $\phi=0$ of the function $F$.
By making use of the generalized version of the Bernoulli polynomials $B^{(m)}_n(t)$ defined by~\cite{Norlund1924}:
\begin{equation*}
\left(\frac{x}{\e^x-1}\right)^m \e^{tx} =\sum_{n=0}^\infty\frac{B^{(m)}_n(t)}{n!}x^n,
\end{equation*}
the function
\begin{equation*}
F(\phi)=\frac{\sinh(\phi)}{\sinh^{2m+3}(\phi/2)}
    = \frac{2^{2m+2}}{\phi^{2m+3}}\left(\frac{\phi}{\e^\phi-1}\right)^{2m+3}
        (\e^\phi - \e^{-\phi})\,\e^{\phi\tfrac{2m+3}{2}}
\end{equation*}
can be written as a convergent power series about $\phi=0$:
\begin{equation*}
\begin{split}
F(\phi)=&2^{2m+2}\sum_{k=0}^\infty \frac{B_k^{(2m+3)}(m+5/2)
    -B_k^{(2m+3)}(m+1/2)}{k!} \phi^{k-2m-3}\\
=&-2^{2m+3}\sum_{k=0}^\infty\frac{B_{2k+1}^{(2m+3)}(m+1/2)}{(2k+1)!}
    \phi^{2k-2m-2},
\end{split}
\end{equation*}
where in the last step we used the identities:
\begin{equation*}
\begin{split}
B_{2k+1}^{(2m+3)}(m+5/2) + B_{2k+1}^{(2m+3)}(m+1/2) =& 0,\\
B_{2k}^{(2m+3)}(m+5/2) -B_{2k}^{(2m+3)}(m+1/2) =& 0,
\end{split}
\end{equation*}
following from the parity of $F(\phi)$ and of the function $\coth(\phi)F(\phi)$.
The theorem~\ref{th1} can be now applied and we obtain:
\begin{equation}\label{sas}
S_{F,2m+2}(\phi)\sim 
-2^{2m+3}\sum_{k=0}^{m+1} \frac{B_{2k+1}^{(2m+3)}(m+1/2)}{(2k+1)!}\phi^{2k}
(1-2^{1+2k-2m-2})\zeta(2+2m-2k).
\end{equation}
By using the identity
$$
 \zeta(2n)=\frac {(-1)^{n+1}B_{2n}(2\pi )^{2n}}{2(2n)!}=-\frac {(2\pi i)^{2n}B_{2n}}{2(2n)!},
$$
where $B_{2n}$ are Bernoulli numbers, the asymptotic expansion \eqref{sas} can be rewritten as:
\begin{equation}\label{sas2}
\begin{split}
S_{F,2m+2}(\phi)\sim &\!-2^{2m+2}\!\!\sum_{k=0}^{m+1}(2\pi i)^{2m+2-2k}
    \frac{B_{2m+2-2k}}{(2\!+\!2m\!-\!2k)!} \frac{B_{2k+1}^{(2m+3)}(m+\tfrac{1}{2})}{(2k\!+\!1)!}
    (2^{1+2k-2m-2}-1)\phi^{2k}\\
= & -2^{2m+2}\sum_{k=0}^{m+1}(2\pi i)^{2k}
    \frac{B_{2k}}{(2k)!} \frac{B_{2m+3-2k}^{(2m+3)}(m+\tfrac{1}{2})}{(2m+3-2k)!}
    (2^{1-2k}-1)\phi^{2m+2-2k}.
\end{split}
\end{equation}
The generalized Bernoulli polynomials fulfill the following identity (see proof below):
\begin{equation}\label{eq:SumId}
\sum_{k=0}^{m+1}\frac{B_{2k}}{(2k)!}
    \frac{B_{2m+3-2k}^{(2m+3)}(m+\tfrac{1}{2})}{(2m+3-2k)!}(2^{1-2k}-1)=0.
\end{equation}
Using the above identity, after simple calculations, the equation \eqref{sas2} can be
recast as:
\begin{equation*}
\begin{split}
S_{F,2m+2}(\phi)\sim 2^{2m+2}\!\sum_{k=0}^{m}\! \frac{B_{2k+2}}{(2k\!+\!2)!}
    \frac{B_{2m+1-2k}^{(2m+3)}(m\!+\!\tfrac{1}{2})}{(2m+1-2k)!}(2^{-1-2k}\!-\!1)\phi^{2m-2k}
    \left(\phi^{2k+2}\!-\!(2\pi i)^{2k+2} \right).
\end{split}
\end{equation*}
Finally, taking into account the relation between the value of Bernoulli polynomials
and the Bernoulli numbers:
\begin{equation*}
B_{2k}\left(\tfrac{1}{2}\right) =  (2^{1-2k}-1)B_{2k},
\end{equation*}
we are in a position to write down the analytic distillation of the series $S_{F,2m+2}$
in the following form:
\begin{equation}
\label{eq:FermiPoly}
\begin{split}
\dist_0 S_{F,2m+2}(\phi)= & -2^{2m+2}\sum_{k=0}^{m+1}(2\pi i)^{2k}
    \frac{B_{2k}\left(\tfrac{1}{2}\right)}{(2k)!} \frac{B_{2m+3-2k}^{(2m+3)}
    \left(m+\tfrac{1}{2}\right)}{(2m+3-2k)!}\phi^{2m+2-2k}\\
= &2^{2m+2}\sum_{k=0}^{m} \frac{B_{2k+2}\left(\tfrac{1}{2}\right)}{(2k+2)!}
    \frac{B_{2m+1-2k}^{(2m+3)}\left(m+1/2\right)}{(2m+1-2k)!}\phi^{2m-2k}
    \left(\phi^{2k+2}-(2\pi i)^{2k+2} \right).
\end{split}
\end{equation}
The physical values are obtained by setting $\phi = \ii a/T_0$. As the right hand side
of manifestly shows, the polynomials in eq.~\eqref{eq:FermiPoly} vanish for $T_0=a/2\pi$, which 
is just the Unruh temperature. 

\subsection{Proof of the equation \texorpdfstring{\eqref{eq:SumId}}{(B.4)} }

We first show that for any integer $M\geq 0$:
\begin{equation}
\label{eq:ZeroGenBer}
B_{2M+1}^{(2M+2)}(M)=0.
\end{equation}
From the Euler Gamma ratio representation~\cite{LukeVol1}, with $z\in\mathbb{C}$ and two 
integers $\alpha,\beta\geq 0$, we have
\begin{equation}\label{eq:gammaratiorepgen}
\frac{\Gamma(z+\alpha)}{\Gamma(z-\beta)}=\sum_{l=0}^{\alpha+\beta} \frac{(\alpha+\beta)!}{l!}
    \frac{B_{\alpha+\beta-l}^{(1+\alpha+\beta)}(\alpha)}{(\alpha+\beta-l)!} z^{l}.
\end{equation}
Then for $M\geq 0$, setting $\alpha=M$ and $\beta=M+1$, we have from \eqref{eq:gammaratiorepgen}
\begin{equation*}
\begin{split}
\frac{\Gamma(z+M)}{\Gamma(z-M-1)}=
    &\sum_{l=0}^{2M+1} \frac{(2M+1)!}{l!(2M+1-l)!} B_{2M+1-l}^{(2M+2)}(M) z^{l}.
\end{split}
\end{equation*}
At $z=0$, the ratio of Gamma functions in the above expression vanishes because:
\begin{equation*}
\frac{\Gamma(M)}{\Gamma(-M-1)}=\Gamma(M)\Gamma(M+1)\frac{\sin(M\pi)}{\pi}=0.
\end{equation*}
However, for $z=0$ in \eqref{eq:gammaratiorepgen} the series contains only one term:
\begin{equation*}
\frac{\Gamma(M)}{\Gamma(-M-1)}=B^{(2M+2)}_{2M+1}\left(M\right)=0,
\end{equation*}
which proves the equation \eqref{eq:ZeroGenBer}.

We are now in a position to prove the identity~\eqref{eq:SumId}. Consider the following 
series, which follows from the generating function of the Bernoulli numbers:
\begin{equation*}
\left( \frac{x}{\e^x-1}+\frac{x}{2}\right)=\sum_{j=0}^\infty \frac{B_{2j}}{(2j)!} x^{2j},
\end{equation*}
and
\begin{equation*}
2\left( \frac{x/2}{\e^{x/2}-1}+\frac{x}{4}\right)
    =\sum_{j=0}^\infty \frac{B_{2j}}{(2j)!} 2^{1-2j}  x^{2j}.
\end{equation*}
Subtracting them, we obtain:
\begin{equation}
\label{eq:Cauchy1}
\sum_{j=0}^\infty \frac{B_{2j}}{(2j)!} (2^{1-2j}-1) x^{2j}
=2\left( \frac{x/2}{\e^{x/2}-1}+\frac{x}{4}\right)-\left( \frac{x}{\e^x-1}+\frac{x}{2}\right)
=\frac{x\, \e^{\frac{x}{2}} }{\e^x-1}.
\end{equation}
We also need the series:
\begin{equation}\label{eq:Cauchy2}
\sum_{j=0}^\infty \frac{B_j^{(2m+3)}(m+\tfrac12)}{j!} x^j
    =\left(\frac{x\, \e^{\frac{x}{2}} }{\e^x-1}\right)^{2m+3}\e^{-x}.
\end{equation}
The function:
\begin{equation*}
h(x)=\left(\frac{x\, \e^{\frac{x}{2}} }{\e^x-1}\right)
    \left(\frac{x\, \e^{\frac{x}{2}} }{\e^x-1}\right)^{2m+3}\e^{-x}
    =\left(\frac{x\, \e^{\frac{x}{2}} }{\e^x-1}\right)^{2m+4}\e^{-x},
\end{equation*}
has the following power series representation:
\begin{equation}\label{eq:fseries}
h(x)= \sum_{n=0}^\infty a_n x^n
= \sum_{n=0}^\infty \frac{B_n^{(2m+4)}(m+1)}{n!} x^n.
\end{equation}
It also possible to find another series representation of the function $h$ by making
the Cauchy product of the series in \eqref{eq:Cauchy1} and \eqref{eq:Cauchy2}:
\begin{equation}
\label{eq:fCauchy}
\begin{split}
h(x) = & \left[\frac{x\, \e^{\frac{x}{2}} }{\e^x-1}\right]
    \left[\left(\frac{x\, \e^{\frac{x}{2}} }{\e^x-1}\right)^{2m+3}\e^{-x}\right]\\
    =&\left[ \sum_{j=0}^\infty \frac{B_{2j}}{(2j)!} (2^{1-2j}-1) x^{2j}\right]
    \left[\sum_{j=0}^\infty \frac{B_j^{(2m+3)}(m+\tfrac12)}{j!} x^j \right]\\
=& \sum_{k=0}^\infty \sum_{l=0}^k \frac{B_{2l}}{(2l)!} (2^{1-2l}-1)
    \frac{B_{k+l-2l}^{(2m+3)}(m+\tfrac12)}{(k+l-2l)!} x^{k+l}.
\end{split}
\end{equation}
Comparing the \eqref{eq:fCauchy} with the \eqref{eq:fseries}, it can be seen that the $(2m+3)$-th coefficient
of the \eqref{eq:fseries} can be retrieved from the \eqref{eq:fCauchy} by setting $k=2m+3-l$ and
summing $l$ from 0 to $m+1$:
\begin{equation*}
a_{2m+3}= \frac{B_{2m+3}^{(2m+4)}(m+1)}{(2m+3)!}= \sum_{l=0}^{m+1} \frac{B_{2l}}{(2l)!} (2^{1-2l}-1)
    \frac{B_{2m+3-2l}^{(2m+3)}(m+\tfrac12)}{(2m+3-2l)!}.
\end{equation*}
Taking into account the eq.~\eqref{eq:ZeroGenBer} with $M=m+1$, it turns out that $a_{2m+3}$ 
vanishes, that is:
\begin{equation*}
\sum_{l=0}^{m+1} \frac{B_{2l}}{(2l)!} (2^{1-2l}-1)
    \frac{B_{2m+3-2l}^{(2m+3)}(m+\tfrac12)}{(2m+3-2l)!}=0,
\end{equation*}
which is the identity~\eqref{eq:SumId} we wanted to prove.

\section{Analytic distillation for pure rotation}
\label{series rotation}

We go through some steps in the calculation of the mean value of the stress-energy tensor 
and the axial current of the massless Dirac field at global equilibrium with rotation. 
In this case, the tensors $\Lambda^n$, $S(\Lambda)^n$, $A$ and $A_5$ in eq.~\eqref{definition A} and $A_5$ in 
eq.~\eqref{definition A5} are obtained from the eq.~\eqref{vorticity rotation} and are given in \eqref{eq:RotMatrices}.

Plugging these expression into the eq.~\eqref{stress energy tensor formula with betatilde}
and~\eqref{eq:AxialMassless}, one obtains the exact mean value of the canonical
stress-energy tensor and of the axial current as a series. The Belinfante stress-energy tensor 
is then obtained form the canonical one from the simple relation \eqref{belinfante}:
\begin{equation*}
T_B^{\mu\nu}(x)_I =\lim_{B\rightarrow\phi}\frac{1}{2\pi^2}\sum_{n=1}^{\infty}\frac{(-1)^{n+1}}
{\left(n^2\phi^2+4B^2r^2T_0^2\sin^2\left(\frac{n\phi}{2}\right)\right)^3}\Theta_n^{\mu\nu}(x),
\end{equation*}
where the tensor $\Theta_n(x)$ has the following components:
\begin{align*}
    &\Theta^{00}_n(x)=8B^4T_0^4\cos\left(\frac{n\phi}{2}\right)[3n^2\phi^2+2B^2r^2T_0^2(\cos n\phi-1)],\\
    &\Theta^{11}_n(x)=8 B^4 T_0^4 \cos \left(\frac{n\phi }{2}\right) \left(B^2 T_0^2 \left(r^2-4 y^2\right)+n^2\phi ^2\right)-8 B^6 T_0^6
         \left(r^2-4 y^2\right) \cos \left(\frac{3 n\phi }{2}\right),\\
    &\Theta^{22}_n(x)=8 B^4 T_0^4 \cos \left(\frac{n\phi }{2}\right) \left(B^2 T_0^2 \left(r^2-4 x^2\right)+n^2\phi ^2\right)-8 B^6 T_0^6
        \left(r^2-4 x^2\right) \cos \left(\frac{3 n\phi }{2}\right),\\
    &\Theta^{33}_n(x)=8 B^4 T_0^4 \cos \left(\frac{n\phi }{2}\right)\left(n^2\phi^2+4B^2r^2T_0^2\sin^2\left(\frac{n\phi}{2}\right)\right),\\
    &\Theta^{01}_n(x)=16 i B^5 T_0^5 y n\phi  \sin \left(\frac{n\phi }{2}\right) (\cos (n\phi )+3),\\
    &\Theta^{02}_n(x)=-16 i B^5 T_0^5 x n\phi  \sin \left(\frac{n\phi }{2}\right) (\cos (n\phi )+3),\\
    &\Theta^{12}_n(x)=64 B^6 T_0^6 x y \sin \left(\frac{n\phi }{2}\right) \sin (n\phi ).
\end{align*}
Note the introduction of the auxiliary real parameter $B$ and the exchange of the limit 
and series as discussed in subsection~\ref{sec:Rot}, with $\phi$ and $B$ both real.

The calculation proceeds as outlined in subsection~\ref{sec:Rot}: by using theorem \ref{th1}, 
the power asymptotic series in $\phi$ of the $B$-dependent components of the stress-energy 
tensor are determined; then, the limit $B \to \phi$ is taken and the resulting expressions 
implicitly provide the full asymptotic power series in $\phi$ about $\phi=0$ which are needed
to determine the analytic distillate. Finally, the vorticity is continued to its physical value 
via the mapping $\phi\rightarrow \ii\omega/T_0$. Eventually, the following expressions are obtained:
\begin{align*}
T^{00}_B=&\frac{7}{180} \pi ^2 \left(4 \gamma^2-1\right) \gamma ^4 T_0^4+\frac{1}{72} 
\left(24 \gamma ^4-16 \gamma ^2+1\right) \gamma ^4 T_0^2 \omega ^2+\\
&+\frac{\left(960 \gamma ^6-1128 \gamma ^4+196 \gamma ^2+17\right) \gamma ^4 \omega ^4}{2880 \pi ^2},\\ 
T^{11}_B=&\frac{7}{180} \pi ^2 \gamma ^4 T_0^4+\frac{1}{360} \gamma ^4 T_0^2 \omega ^2
\left(4 \gamma ^2 \left(14\pi ^2 T_0^2 y^2+5\right)-5\right)+,\\&+\frac{\gamma ^4 \omega ^4 
\left(120 \gamma ^4 \left(8 \pi ^2 T_0^2 y^2+1\right)-8 \gamma ^2\left(20 \pi ^2 T_0^2 y^2+11\right)-17\right)}
{2880 \pi ^2}+\\
&+\frac{\gamma ^6 \left(240 \gamma ^4-132 \gamma ^2-17\right) y^2 \omega ^6}{720 \pi ^2},\\
T^{22}_B&=\frac{7}{180} \pi ^2 \gamma ^4 T_0^4+\frac{1}{360} \gamma ^4 T_0^2 \omega ^2 
\left(4 \gamma ^2 \left(14\pi ^2 T_0^2 x^2+5\right)-5\right)+,\\&+\frac{\gamma ^4 \omega ^4 
\left(120 \gamma ^4 \left(8 \pi ^2 T_0^2 x^2+1\right)-8 \gamma ^2\left(20 \pi ^2 T_0^2 x^2+11\right)-17\right)}
{2880 \pi ^2}+\\
&+\frac{\gamma ^6 \left(240 \gamma ^4-132 \gamma ^2-17\right) x^2 \omega ^6}{720 \pi ^2}
\end{align*}
\begin{align*}
T^{33}_B&=\frac{7}{180} \pi ^2 \gamma ^4 T_0^4+\frac{1}{72}\left(4 \gamma ^2-1\right) \gamma ^4 T_0^2 \omega ^2+\frac{\left(120 \gamma ^4-88 \gamma ^2-17\right) \gamma ^4 \omega ^4}{2880 \pi ^2},\\
T^{01}_B&=-\frac{7}{45} \pi ^2 \gamma ^6 T_0^4 y \omega +\frac{1}{9} \left(1-3 \gamma ^2\right) 
\gamma ^6 T_0^2 y \omega ^3-\frac{\left(80\gamma ^4-64 \gamma ^2+15\right) \gamma ^6 y \omega ^5}{240 \pi ^2}\\
T^{02}_B&=\frac{7}{45} \pi ^2 \gamma ^6 T_0^4 x \omega -\frac{1}{9} \left(1-3 \gamma ^2\right) \gamma ^6 
T_0^2 x \omega ^3+\frac{\left(80\gamma ^4-64 \gamma ^2+15\right)\gamma ^6 x \omega ^5}{240 \pi ^2},\\
T^{12}_B&=-\frac{7}{45} \pi ^2 \gamma ^6 T_0^4 x y \omega ^2+\frac{1}{18} \left(1-6 \gamma ^2\right) 
\gamma ^6 T_0^2 x y \omega^4-\frac{\left(240 \gamma ^4-132 \gamma ^2-17\right) \gamma ^6 x y \omega ^6}{720 \pi ^2},
\end{align*}
where $\gamma=1/\sqrt{1-r^2\omega^2}$. To finally derive the coefficients appearing in the 
decomposition of eq.~\eqref{general stress energy tesor}, the relations:
\begin{equation*}
\begin{split}
    u^\mu=&\gamma  \left(1,-y \omega ,x \omega ,0\right),\quad
    \alpha^\mu= \left(0,-\frac{\gamma  x \omega ^2}{T_0},-\frac{\gamma  y \omega^2}{T_0},0\right),\\
    w^\mu=& \left(0,0,0,\frac{\gamma  \omega }{T_0}\right),\quad
    l^\mu= \left(\frac{\gamma  \left(\gamma^2-1\right) \omega^2}{T_0^2},-\frac{\gamma^3 y \omega ^3}{T_0^2},\frac{\gamma^3 x \omega^3}{T_0^2},0\right),
\end{split}
\end{equation*}
and:
\begin{equation*}
\beta^2(x)=\frac{1}{T_0^2\gamma^2},\quad \alpha^2(x)=-(\gamma^2-1)\frac{\omega^2}{T_0^2},
\quad w^2(x)=-\gamma^2\frac{\omega^2}{T_0^2},
\end{equation*}
must be used. The results are quoted in \eqref{coeffrot}.

The mean value of the axial current is obtained likewise. For equilibrium with rotation, 
the only non-vanishing component of the series~~\eqref{eq:AxialMassless} is directed along 
the axis of rotation:
\begin{equation*}
 j_A^3(x)_I=\lim_{B\rightarrow \phi}- \frac{4\ii B^3T_0^3}{\pi^2}\sum_{n=1}^{\infty}(-1)^{n+1}
 \frac{n\phi \sin\left(\frac{n\phi}{2}\right)}{\left(n^2\phi^2+4B^2T_0^2r^2
 \sin^2\left(\frac{n\phi}{2}\right)\right)^2}.
\end{equation*}
The series is handled like for the stress-energy tensor components and it yields:
\begin{equation*}
j_A^3(x)=\left.\dist_0 j_A^3(x)_I\right|_{\phi=\ii\omega/T_0}
=\frac{1}{6}\gamma^4T_0^2\omega+\frac{\gamma^4\omega^3}{24\pi^2}(4\gamma^2-3).
\end{equation*}
This is the same result obtained in \cite{ambrus2019exact}. Employing the tetrad defined above, 
the axial current can be written in the form of eq.~\eqref{eq:JARot}.

\section{Analytic distillation for rotation and acceleration}
\label{series acc-rot}

Here we go through some steps of the analytic distillation of the stress-energy tensor and 
the axial current of the free massless Dirac field when equilibrium includes both a rotation 
and acceleration along the $z$-axis; the corresponding calculation for the scalar field was 
carried out in ref.~\cite{ScalarField}.
The series of the canonical/Belinfante stress-energy tensor and the axial current are obtained by using
the expressions~\eqref{eq:matricesRotAcc} in eqs.~\eqref{stress energy tensor formula with betatilde} 
and~\eqref{eq:AxialMassless}.

In order to apply theorem \ref{th1}, the series obtained with eq.~\eqref{stress energy tensor formula with betatilde} 
and~\eqref{eq:AxialMassless} are to be rewritten in a suitable form by introducing auxiliary 
parameters (thereafter $B$, $C$ and, when needed $D$). Particularly, the Belinfante stress-energy 
tensor can be written as:
\begin{equation*}
    T^{\mu\nu}_B(x)_I=\lim_{B,C,D\rightarrow\bar{B},\bar{C},\bar{D}}\frac{1}{2\pi^2}
    \sum_{n=1}^\infty\frac{(-1)^{n+1}}{2 \left(B \sinh ^2\left(\frac{n \Phi }{2}\right)
    +C \sin ^2\left(\frac{n \phi }{2}\right)\right)^3}\Theta_n^{\mu\nu}(x,B,C,D),
\end{equation*}
where ($t,z$ being the Cartesian coordinates):
\begin{align}\label{bcd1}
    \bar{B}=\Phi^2 T_0^2 t^2 + (1 - i \Phi T_0 z)^2, &&\bar{C}=r^2\Phi^2T_0^2,
\end{align}
and $\bar{D}$ is a component-dependent quantity:
\begin{equation}\label{bcd2}
\bar{D}_{00}=(T_0 \Phi  z+i),\quad
\bar{D}_{11}=T_0 \Phi y,\quad 
\bar{D}_{22}=T_0 \Phi x,\quad
\bar{D}_{33}=T_0 \Phi t.
\end{equation}
In the above expressions, limits can be exchanged with the series because they are
uniformly convergent series of continuous functions of the arguments $B,C,D$, for 
real values and for real $\phi$ and $\Phi$ \footnote{Indeed the series does not converge for 
$B=0$, which is impossible as long as $T_0 > 0.$}. The tensor $\Theta$ reads:
\begin{align*}
\Theta^{00}_n&=-T_0^4 \Phi^4 \cosh(\tfrac{n\Phi}{2}) \cos(\tfrac{n\phi}{2})
\left( B \sinh^2(\tfrac{n\Phi}{2}) +C \sin^2(\tfrac{n\phi}{2})+2 D_{00}^2 (\cosh (n \Phi )-1)
\right),\\
\Theta_n^{11}&=T_0^4 \Phi^4 \cosh(\tfrac{n\Phi}{2}) \cos(\tfrac{n\phi}{2})
\left(B \sinh^2(\tfrac{n \Phi}{2}) +C \sin ^2(\tfrac{n\phi}{2})+2D^2_{11} 
(\cos (n \phi )-1)\right),\\
\Theta_n^{33}&=T_0^4 \Phi ^4 \cosh (\tfrac{n\Phi}{2}) \cos (\tfrac{n\phi}{2})
\left(B \sinh ^2(\tfrac{n \Phi}{2})+C \sin ^2(\tfrac{n\phi}{2})-2 D^2_{33} (\cosh (n \Phi )-1)\right),\\
\Theta_n^{01}&=T_0^5 \Phi ^5 y (i+T_0 \Phi  z) \sinh (\tfrac{n\Phi}{2}) \sin (\tfrac{n\phi}{2})  
(\cosh (n \Phi )+\cos (n \phi )+2),\\
\Theta_n^{03}&=- T_0^5 \Phi ^5 t(i+T_0
   \Phi  z)\sinh (\tfrac{n\Phi}{2}) \sinh (n \Phi ) \sin (n \phi ) \csc (\tfrac{n\phi}{2}),\\
\Theta_n^{12}&=T_0^6 \Phi ^6 x y \sinh (n \Phi ) \text{csch}(\tfrac{n\Phi}{2}) 
\sin (\tfrac{n\phi}{2}) \sin (n \phi )\\
   \Theta_n^{13}&= T_0^6 \Phi ^6 ty \sinh (\tfrac{n\Phi}{2}) \sin (\tfrac{n\phi}{2}) 
   (\cosh (n \Phi )+\cos (n \phi )+2),\\
\Theta_n^{22}&= \Theta_n^{11}(y\mapsto x),\quad
\Theta_n^{02}=\Theta_n^{01}(y\mapsto-x),\quad
\Theta_n^{23}=\Theta_n^{12}(y\mapsto-x).
\end{align*}
As they stand, the components of the stress-energy tensor are proportional to alternating-sign 
series of functions of $n\phi$ and $n\Phi$. To use the theorem \ref{th1}, we define the 
map:
\be\label{polar transf}
\Phi=\xi \cos\theta  \qquad \qquad \phi=\xi\sin\theta
\ee
thereby obtaining series of functions of $n\xi$. The asymptotic expansion of each component 
of the stress-energy tensor can now be calculated. Proceeding as in the previous cases, i.e.
evaluating the limits in the auxiliary parameters, obtaining the analytic distillate and
continuing to the physical values of thermal vorticity, we get the exact TEVs. 
To sketch the various steps, we focus on the time-time component of the Belinfante stress-energy
tensor as an example:
\begin{equation*}
\begin{split}
T^{00}_B(x)_I=&-\frac{T_0^4 \Phi^4}{4\pi^2}\sum_{n=1}^\infty\lim_{B,C,D\rightarrow\bar{B},\bar{C},\bar{D}}(-1)^{n+1}
\cosh \left(\frac{n \Phi }{2}\right) \cos \left(\frac{n \phi }{2}\right)\times \\
&\frac{B \sinh ^2\left(\frac{n \Phi
   }{2}\right)+C \sin ^2\left(\frac{n \phi }{2}\right)+2 (\cosh (n \Phi )-1) D^2}{\left(B \sinh
   ^2\left(\frac{n \Phi }{2}\right)+C \sin ^2\left(\frac{n \phi }{2}\right)\right)^3},
\end{split}
\end{equation*}
where $\bar{B}, \bar{C}, \bar{D}$ like in eq.~\eqref{bcd1}\eqref{bcd2} ($D=D_{00}$ and $\bar{D}=\bar{D}_{00}$ 
are implied). The transformation \eqref{polar transf}
is applied and a series of functions $f(n\xi)$ is thus obtained:
\begin{equation*}
\begin{split}
T^{00}_B(x)_I=&\lim_{B,C,D\rightarrow\bar{B},\bar{C},\bar{D}}
-\frac{T_0^4 \xi^4\cos^4\theta}{4\pi^2}\sum_{n=1}^\infty
\frac{(-1)^{n+1}\cosh \left(\tfrac{n \xi\cos\theta }{2}\right) 
\cos \left(\tfrac{n \xi\sin\theta}{2}\right)}{\left(B \sinh^2\left(\tfrac{n \xi\cos\theta }{2}\right)
+C \sin ^2\left(\tfrac{n \xi\sin\theta }{2}\right)\right)^3}\times\\
&\left(B \sinh^2\left(\tfrac{n \xi\cos\theta}{2}\right)
+C \sin^2\left(\tfrac{n \xi\sin\theta }{2}\right)+2 (\cosh(n \xi\cos\theta )-1) D^2\right)\\
\equiv &\lim_{B,C,D\rightarrow\bar{B},\bar{C},\bar{D}}-\frac{T_0^4 \xi^4\cos^4\theta}{4\pi^2} 
G(\theta,\xi,B,C,D).
\end{split}
\end{equation*}
The asymptotic power expansion in $\xi$ of the series $G$ can now be obtained by using theorem~\ref{th1}:
\begin{equation*}
\begin{split}
G \sim&
\frac{7 \pi^4 \left(\left(B+4 D^2\right) c^2+C s^2\right)}{45 \xi^4 \left(B c^2+C s^2\right)^3}\\
&-\frac{\pi^2}{18 \xi^2 \left(B c^2 +C s^2\right)^4}\left[s^2 c^4 \left(3 B^2-2 B \left(C-6 D^2\right)
    -20 C D^2\right)\right.\\ 
&\left.+C s^4 c^2 \left(2 B-3 \left(C+4 D^2\right)\right)
    +B\left(B+4 D^2\right) c^6-C^2 s^6\right]\\
&+\frac{1}{720\left(B c^2+C s^2\right)^5}\left[B s^2 c^8 \left(30 B^2-139 B C+120
       B D^2-664 C D^2\right)\right.\\
&+C s^6 c^4 \left(15 C \left(C+40 D^2\right)-107 B^2-6 B\left(25 C+132 D^2\right)
       \right)\\
&-17 B^2 \left(B+4D^2\right) c^{10}+s^4 c^6 \left(15 B^3
    -30 B^2 \left(5 C-2 D^2\right)\right.\\
&\left.-B C \left(107 C+1200 D^2\right)+364 C^2 D^2\right)\\
&\left.+C^2 s^8 c^2 \left(-139 B+30 C+108 D^2\right)-17 C^3 s^{10}\right],
\end{split}
\end{equation*}
where
\begin{equation*}
c=\cos\theta,\qquad s=\sin\theta.
\end{equation*}
Having obtained the asymptotic power expansion of $G$ in powers of $\xi$ about $\xi=0$, which 
we shall denote as $G_A$, we can compute the limits in the auxiliary parameters and transform the 
variables $\xi, \theta$ back to $\Phi, \phi$. Hence:
\begin{equation*}
    \dist_0 T^{00}_B(x)(\phi,\Phi)_I = \dist_0 \frac{T_0^4\Phi^4}{4\pi^2} G_A(\Phi,\phi,\bar{B},\bar{C},\bar{D}).
\end{equation*}
Finally, we can set $\Phi\rightarrow ia/T_0$ and $\phi\rightarrow i\omega/T_0$ to get 
the physical value by analytic continuation:
\begin{equation*}
    T^{00}_B(x) = \frac{a^4}{4\pi^2} G_A(ia/T_0,i\omega/T_0,\bar{B},\bar{C},\bar{D}),
\end{equation*}
where the arguments of $\bar B, \bar C, \bar D$ in eqs.~\eqref{bcd1}, \eqref{bcd2} are also continued.
The other components can be worked out in a similar fashion. The resulting coefficients of the 
general decomposition \eqref{general stress energy tesor} are quoted in eq.~\eqref{coeffaccrot}. 
They can be extracted by comparing the various components with the decomposition onto
the tetrad:
\begin{align*}
    u^\mu=& \gamma\left(1+az,-\omega y,\omega x,at\right),\\
    \alpha^\mu =& \frac{\gamma}{T_0} \left(a^2 t,-\omega^2x,-\omega^2y,a(1+az)\right),\\
    w^\mu=&\frac{\gamma \omega}{T_0}\left(at,ax,ay,1+az\right),\\
    l^\mu=&\frac{\gamma^3 \omega(a^2+\omega^2)}{T_0^2}\left(\omega r^2(1+az),-y((1+az)^2-a^2t^2),
    x((1+az)^2-a^2t^2),r^2at\right),
\end{align*}
with:
\begin{align*}
    &\beta^2=\frac{a^2 \left(z^2-t^2\right)+2 a z-r^2 \omega ^2+1}{T_0^2}, &&\alpha^2
    =-\frac{a^4 \left(z^2-t^2\right)+2 a^3 z+a^2+r^2 \omega ^4}{T_0^4 \beta^2},\\
    &w^2=-\frac{\omega ^2 \left(a^2 \left(r^2-t^2+z^2\right)+2 a z+1\right)}{T_0^4 \beta^2}, 
    &&\alpha\cdot w=- a\omega/T_0^2,
\end{align*}
where $\gamma=\left((1+az)^2-a^2t^2-\omega^2r^2\right)^{-1/2}$.

The axial current can be obtained from the series~\eqref{eq:AxialMassless}:
\begin{equation*}
    j^\mu_A(x)_I=\lim_{B,C\rightarrow \bar{B},\bar{C}}\frac{\Phi^3T_0^3}{2\pi^2}\sum_{n=1}^\infty \frac{(-1)^{n+1}}{\left(B\sinh\left(\frac{n\Phi}{2}\right)+C\sin\left(\frac{n\phi}{2}\right)\right)^2}\Upsilon^\mu_n
\end{equation*}
where $\bar{B}$ and $\bar{C}$ are the same as in \eqref{bcd1},\eqref{bcd2} and the vector 
$\Upsilon$ reads:
\begin{equation*}
    \begin{split}
        \Upsilon^0&=-\Phi tT_0\sin\left(\frac{n\phi}{2}\right)\sinh\left(\frac{n\Phi}{2}\right),\\
        \Upsilon^3&=-(i+\Phi T_0 z)\sin\left(\frac{n\phi}{2}\right)\sinh\left(\frac{n\Phi}{2}\right) ,\\
        \Upsilon^1&=\Upsilon^0(t\mapsto x),\\
        \Upsilon^2&=\Upsilon^0(t\mapsto y).
    \end{split}
\end{equation*}
The method is the same as for the stress-energy tensor, which eventually yields the covariant expression 
\eqref{eq:JARot}, with no extra dependence on $\alpha \cdot w$.

\bibliographystyle{ieeetr}
\bibliography{biblio}
\clearpage

\end{document}